\documentclass[aps,pre,repreprint,superscriptaddress,twocolumn,showpacs]{revtex4-1}
\usepackage{graphicx}
\usepackage{subfigure}
\begin{document}

% Use the \preprint command to place your local institutional report
% number in the upper righthand corner of the title page in preprint mode.
% Multiple \preprint commands are allowed.
% Use the 'preprintnumbers' class option to override journal defaults
% to display numbers if necessary
%\preprint{}

%Title of paper
\title{Pedestrian dynamics in single-file movement of crowd with different age compositions}

\author{Shuchao Cao}
\affiliation{State Key Laboratory of Fire Science, University of Science and Technology of China, Hefei, 230027, China}
\affiliation{Forschungszentrum J\"ulich GmbH, Institute for Advanced Simulation, 52425 J\"ulich, Germany}
\author{Jun Zhang}
\email[Corresponding author:]{ju.zhang@fz-juelich.de}
\affiliation{Forschungszentrum J\"ulich GmbH, Institute for Advanced Simulation, 52425 J\"ulich, Germany}
\author{Daniel Salden}
\affiliation{Forschungszentrum J\"ulich GmbH, Institute for Advanced Simulation, 52425 J\"ulich, Germany}
\author{Jian Ma}
\affiliation{School of Transportation and Logistics, Southwest Jiaotong University, Chengdu, 610031, China}
\author{Chang'an Shi}
\affiliation{Tianshui Architectural Design Institute, Tianshui, 741000, China}
\author{Ruifang Zhang}
\affiliation{Tianshui Health School, Tianshui, 741000, China}

%\date{\today}

\begin{abstract}

An aging population is bringing new challenges to the management of escape routes and facility design in many countries. This paper investigates pedestrian movement properties of crowd with different age compositions. Three pedestrian groups are considered: young student group, old people group and mixed group. It is found that traffic jams occur more frequently in mixed group due to the great differences of mobilities and self-adaptive abilities among pedestrians. The jams propagate backward with a velocity 0.4 $m/s$ for global density $\rho_g \approx 1.75~m^{-1}$ and 0.3 $m/s$ for $\rho_g > 2.3~m^{-1}$. The fundamental diagrams of the three groups are obviously different from each other and cannot be unified into one diagram by direct non-dimensionalization. Unlike previous studies, three linear regimes in mixed group but only two regimes in young student group are observed in the headway-velocity relation, which is also verified in the fundamental diagram. Different ages and mobilities of pedestrians in a crowd cause the heterogeneity of system and influence the properties of pedestrian dynamics significantly. It indicates that the density is not the only factor leading to jams in pedestrian traffic. The composition of crowd has to be considered in understanding pedestrian dynamics and facility design.

\end{abstract}

% insert suggested PACS numbers in braces on next line
%\pacs{89.75.Fb, 89.65.−s, 89.40.Bb, 89.90.+n}
% insert suggested keywords - APS authors don't need to do this

%\maketitle must follow title, authors, abstract, \pacs, and \keywords
\maketitle

% body of paper here - Use proper section commands
% References should be done using the \cite, \ref, and \label commands
\section{Introduction}

With the increasing size and frequency of mass events, the study on pedestrian dynamics has become an important area in recent years. It is important for designing pedestrian facilities concerning safety and comfort. Meanwhile, it can also provide technique support for effective management of crowd. However, pedestrian crowd is a complex system containing interesting self-organization phenomena and collective effects. The crowd movement is affected by several factors including pedestrian motivation, facility geometry, culture differences etc.. Large discrepancies on pedestrian movement properties, for example the basic fundamental diagram, have been observed from previous studies \cite{Seyfried2010, Zhang2012}.

Besides, aging population, increasing obesity and more people with mobility impairments are bringing new challenges to the management of routine and people movement in many countries. The movement properties of pedestrian crowds with different age compositions are crucial in pedestrian dynamics studies. Up to now, some models considering the movement characteristics of different age groups have been built. The study subjects include elementary student \cite{Tang2015}, elder people \cite{Galiza2011,Spearpoint2012} and disabled residents \cite{Manley2011, Koo2012, Koo2013}. Besides, some pedestrian experiments considering age of humans are also performed \cite{Kholshchevnikov2012, Cuesta2015, Kuligowski2013, Shimura2014}. However, these studies only concentrate on single pedestrian group (children, older adults or disabled person), and the mixed situations are rare to be considered. Furthermore, the existing experiments were carried out with different focuses under different conditions. In this case, it is difficult to compare the movement properties of pedestrians with various age compositions. To reduce the influences from environment and experiment setup, we study the characteristics of pedestrian flow through a series of single-file pedestrian experiments, which involves purely longitudinal interactions among pedestrians and avoids lateral ones. Indeed, similar experiments have already been done in different countries with young test people in the past. Seyfried et al.\cite{Seyfried2005} measured the fundamental diagram for densities up to 2 $m^{-1}$. Chattaraj et al. \cite{Chattaraj2009} in India and Liu et al. \cite{Liu2009} in China investigate the culture difference on the fundamental diagram with university students. In France Jeli\'c et al. \cite{Jelic2012a, Jelic2012b} conducted experiment inside a ring formed by inner and outer round walls to study the properties of pedestrians moving in line. The density in their experiment reached 3 $m^{-1}$ and the stepping behavior and fundamental diagrams were studied. In that study, three regimes (free regime, weakly constrained regime and strongly constrained regime) were distinguished by analyzing the velocity-spatial headway relationship, which is different from previous works \cite{Seyfried2005, Chattaraj2009}. The transitions between these regimes occur at spatial headways of about 1.1 $m$ and 3 $m$ respectively. The adaptation time takes only three discrete values, corresponding to the three regimes. Stop-and-go wave as a special phenomenon of congested flow was also observed from the experiment with soldiers \cite{Seyfried2010b}. However, university students who have small differences in age and mobility were used as test persons in all of these single-file experiments. It is not known whether the same results can be obtained for the younger or older pedestrian groups. Particularly, the dynamics of mixed crowd composed of pedestrians with large age difference has never been investigated. The effect of the induced heterogeneity on the movement properties is necessary to be considered to get deeper and more comprehensive understanding on pedestrian dynamics.

Based on these considerations, in this paper, we perform well-controlled single-file experiments under laboratory condition to investigate the movement properties of crowds with different age compositions. Three pedestrian groups are considered: young student group, old people group and mixed group. The remainder of the paper is organized as follows. In Section \ref{exp-setup}, the setup of the experiment is described in detail. In Section \ref{analysis}, we analyze the characteristics of pedestrian movement and compare the results from different age groups based on trajectories extracted from video recordings. Finally, the conclusions from our investigation are summarized and discussed.

\section{Experiment setup}\label{exp-setup}

\begin{figure}
 \includegraphics[width=0.45\textwidth]{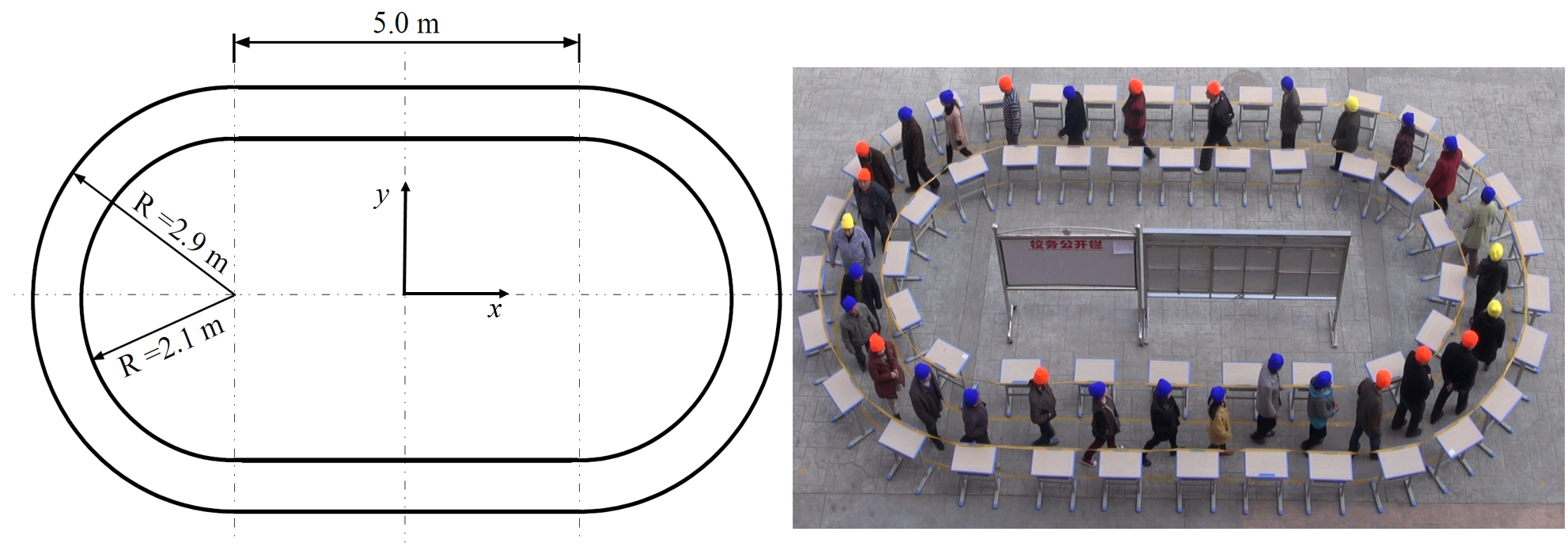}
 \caption{\label{fig-exp-sketch}The schematic illustration of the scenario and a snapshot from the experiment.}
 \end{figure}

The experiment was carried out in 2015 in Tianshui Health School, Gansu, China. FIG. \ref{fig-exp-sketch} displays the sketch of the scenario and a snapshot from the experiment. The closed ring corridor with a central circumference $C$ = 25.7 $m$ includes two 5 $m$ straight parts and two semicircles. The width of corridor was set as 0.8 $m$ to avoid overtaking during movement. The test pedestrians include 80 young students (16 to 18 years old with an average age of 17) from the school and 47 aged adults (45 to 73 years old with an average age of 52) recruited from residential districts nearby. As shown in TABLE \ref{Table_runs}, totally 30 runs were performed with the different numbers and types of participants (15 runs with young student group, 6 runs with old group and other 9 runs with mixed group composed of young student and aged people) in the corridor. It should be noticed that for the mixed group, the students and old adults were arranged in the corridor alternately with a ratio of 1:1 in each run. The participants received the same instruction in each run.  In order to collect enough data especially under steady state and guarantee the repetitiveness and meaningfulness of the results, they were asked to walk in normal way without overtaking for at least 3 minutes. According to the different numbers $N$ of pedestrians in each run, the global densities $\rho_g = N / C$ range from 0.19 to 2.76 $m^{-1}$ for student group, from 0.23 $m^{-1}$ to 1.17 $m^{-1}$ for old adults group and from 0.23 $m^{-1}$ to 2.33 $m^{-1}$ for mixed group. Considering the limited physical strength of aged adults, we have to let part of them rest to guarantee their safety during the experiment. Consequently, the maximum $N$ is only 30 in old group and their movements under very high density situations were not obtained from the experiment.

 % Here is an example of the general form of a table:
% Fill in the caption in the braces of the \caption{} command. Put the label
% that you will use with \ref{} command in the braces of the \label{} command.
% Insert the column specifiers (l, r, c, d, etc.) in the empty braces of the
 %\begin{tabular}{} command.
% The ruledtabular enviroment adds doubled rules to table and sets a
% reasonable default table settings.
% Use the table* environment to get a full-width table in two-column
% Add \usepackage{longtable} and the longtable (or longtable*}
% environment for nicely formatted long tables. Or use the the [H]
% placement option to break a long table (with less control than
% in longtable).
 \begin{table}%[H] add [H] placement to break table across pages
 \caption{\label{Table_runs}The details of the whole experiment.}
 \begin{ruledtabular}
 \begin{tabular}{cccccc}
 &&\multicolumn{2}{c}{Number of pedestrians $[N]$}&&\\[-2mm]
Index	&Name& \multicolumn{2}{c}{\rule{3.8cm}{0.013cm}}&Global density\\
&&Young student&Old adults&$\rho_g$ [$m^{-1}$]&\\
\hline
01	&Young-05	&5	&-&	0.19\\
02	&Young-10	&10	&-&	0.39\\
03	&Young-15	&15	&-&	0.58\\
04	&Young-20	&20	&-	&0.78\\
05	&Young-25	&25	&-&	0.97\\
06	&Young-30	&30	&-	&1.17\\
07	&Young-35	&35	&-&	1.36\\
08	&Young-40	&40	&-	&1.56\\
09	&Young-45	&45	&-	&1.75\\
10	&Young-50	&50	&-	&1.94\\
11	&Young-51	&51	&-	&1.98\\
12	&Young-56	&56	&-	&2.18\\
13	&Young-61	&61	&-	&2.37\\
14	&Young-66	&66	&-	&2.57\\
15	&Young-71	&71	&-&	2.76\\
\hline
01	&Old-06	&-&	6	&0.23\\
02	&Old-11	&-	&11	&0.43\\
03	&Old-16	&-	&16	&0.62\\
04	&Old-21	&-	&21	&0.82\\
05	&Old-26	&-	&26	&1.01\\
06	&Old-30	&-	&30	&1.17\\
\hline
01	&Mixed-06	&3	&3	&0.23\\
02	&Mixed-10	&5	&5	&0.39\\
03	&Mixed-16	&8	&8	&0.62\\
04	&Mixed-20	&10	&10	&0.78\\
05	&Mixed-26	&13	&13	&1.01\\
06	&Mixed-30	&15	&15	&1.17\\
07	&Mixed-36	&18	&18	&1.40\\
08	&Mixed-46	&23	&23	&1.79\\
09	&Mixed-60	&30	&30	&2.33\\
 \end{tabular}
 \end{ruledtabular}
 \end{table}

\begin{figure}
 \centering\subfigure[Young-05]{
 \includegraphics[height=0.125\textheight]{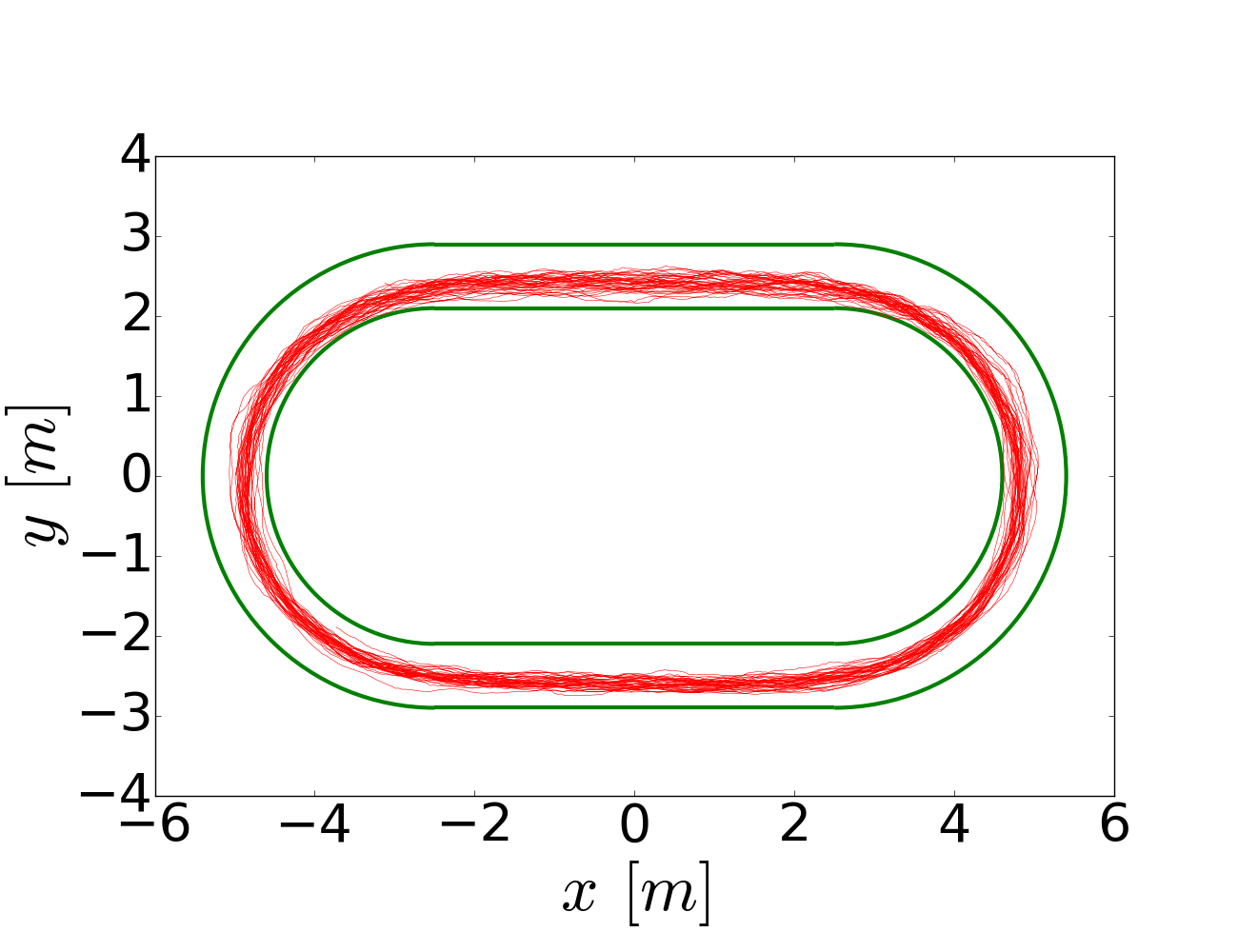}
  \includegraphics[height=0.125\textheight]{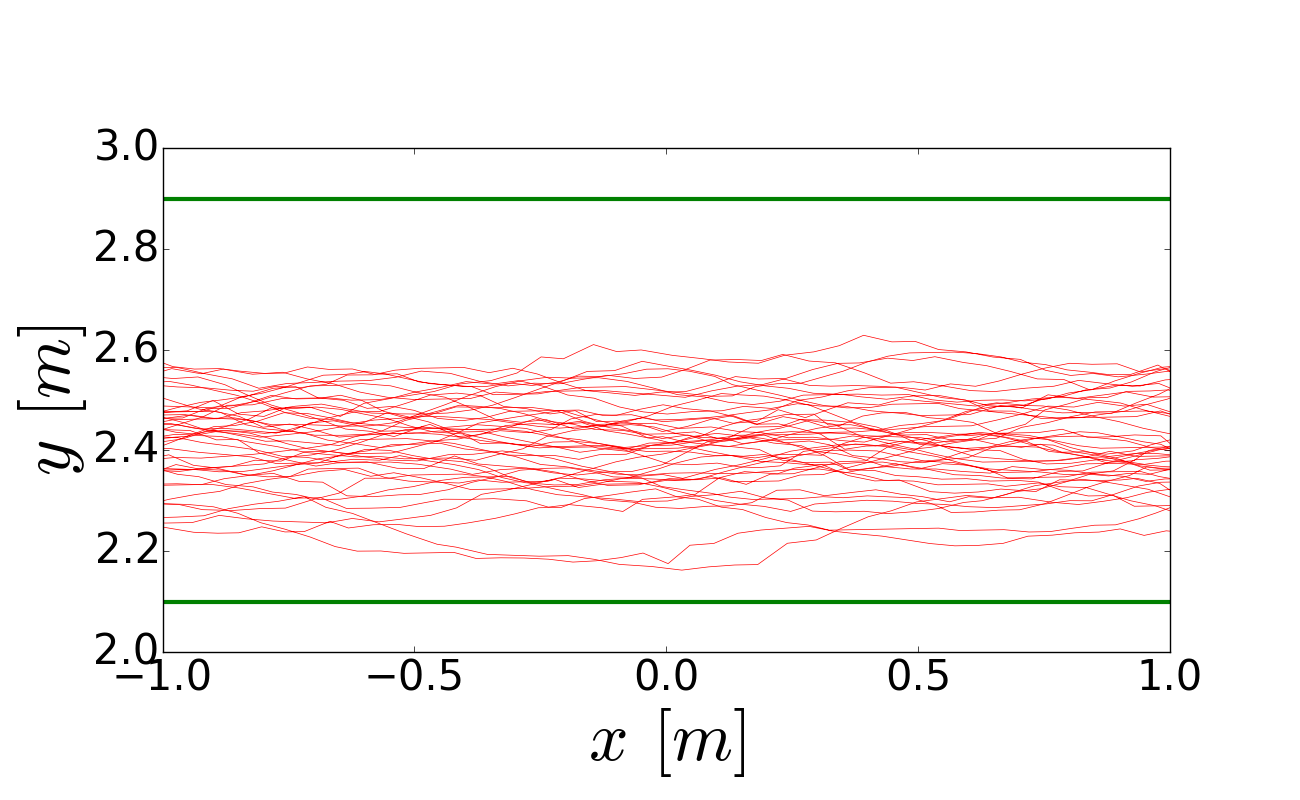}}
 \centering\subfigure[Young-40]{
 \includegraphics[height=0.125\textheight]{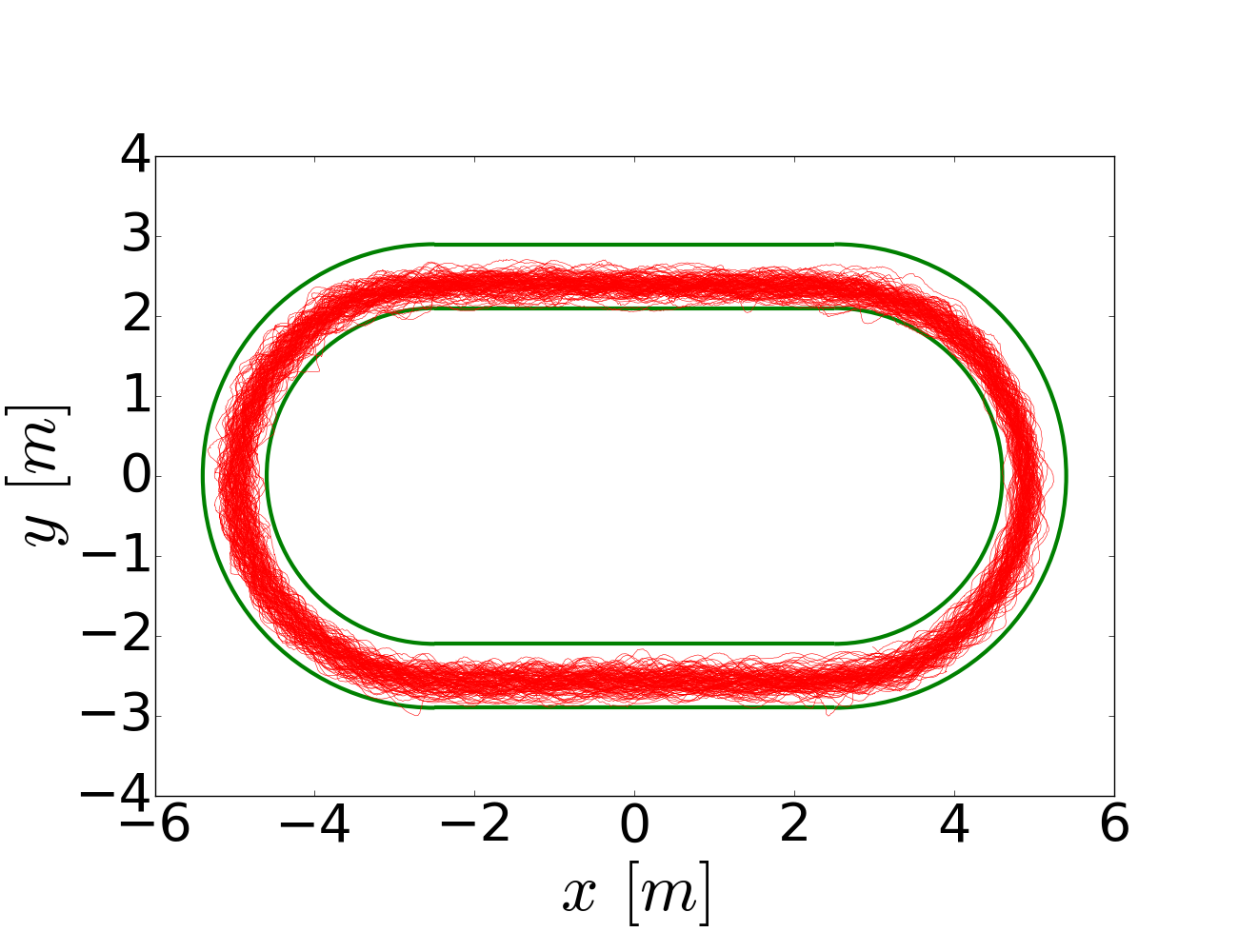}
  \includegraphics[height=0.125\textheight]{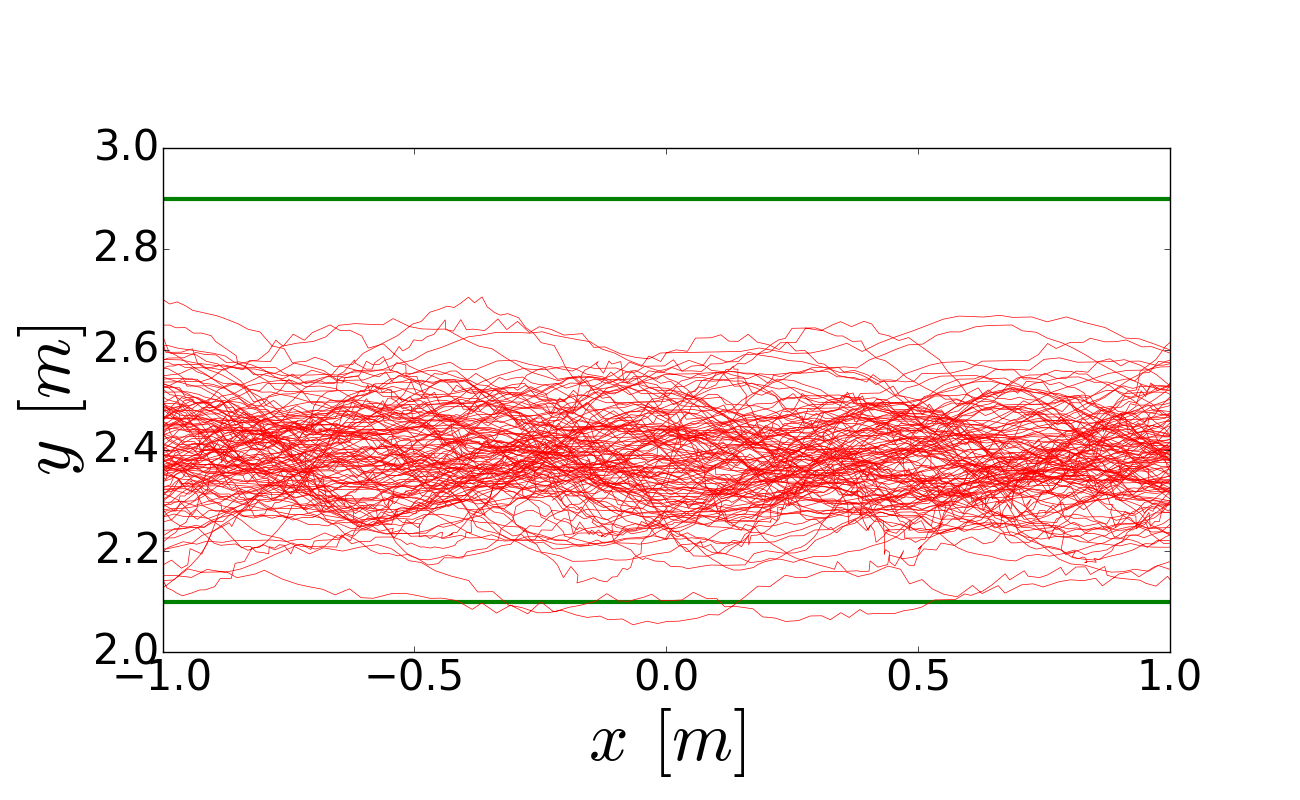}}
 \centering\subfigure[Young-66]{
 \includegraphics[height=0.125\textheight]{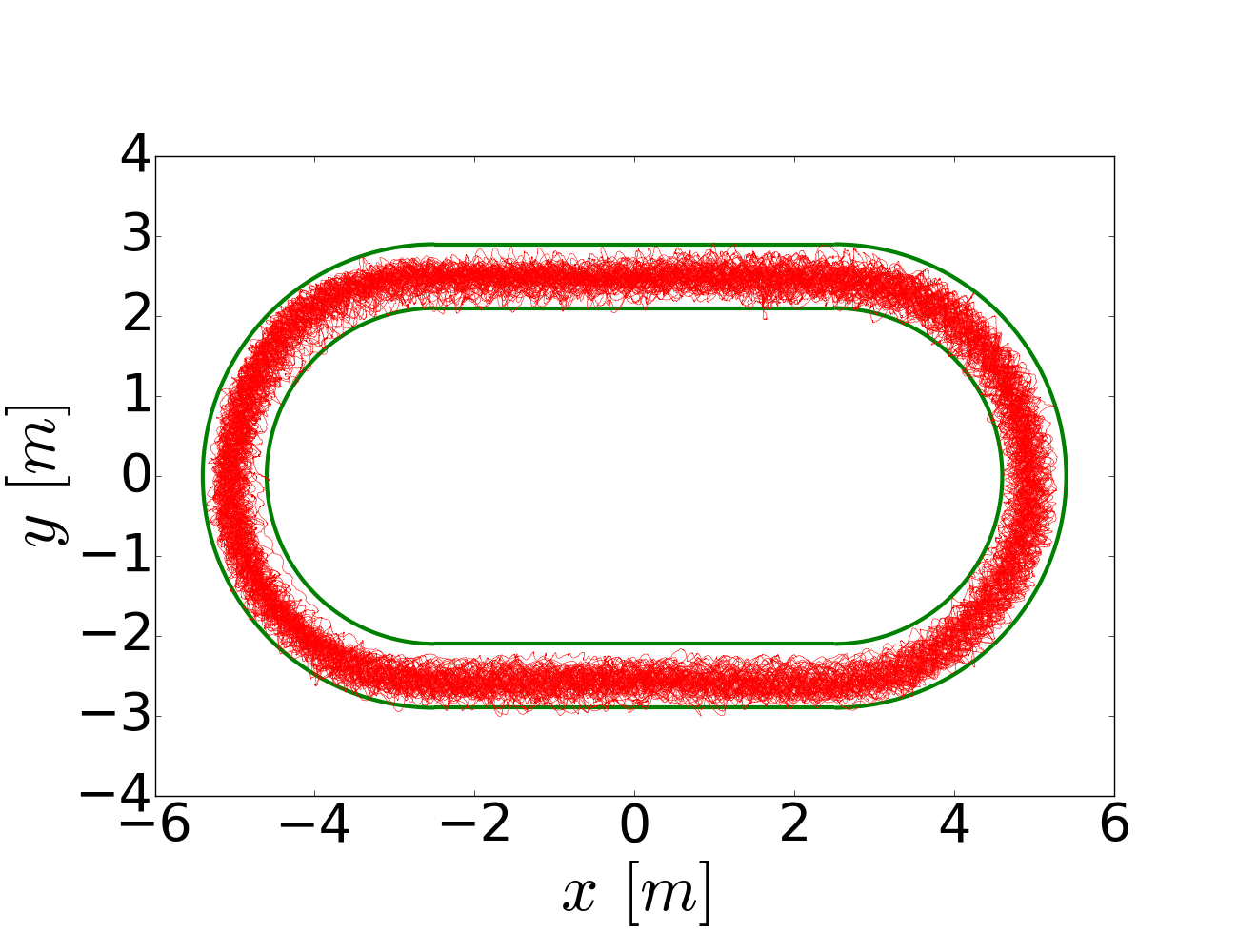}
  \includegraphics[height=0.125\textheight]{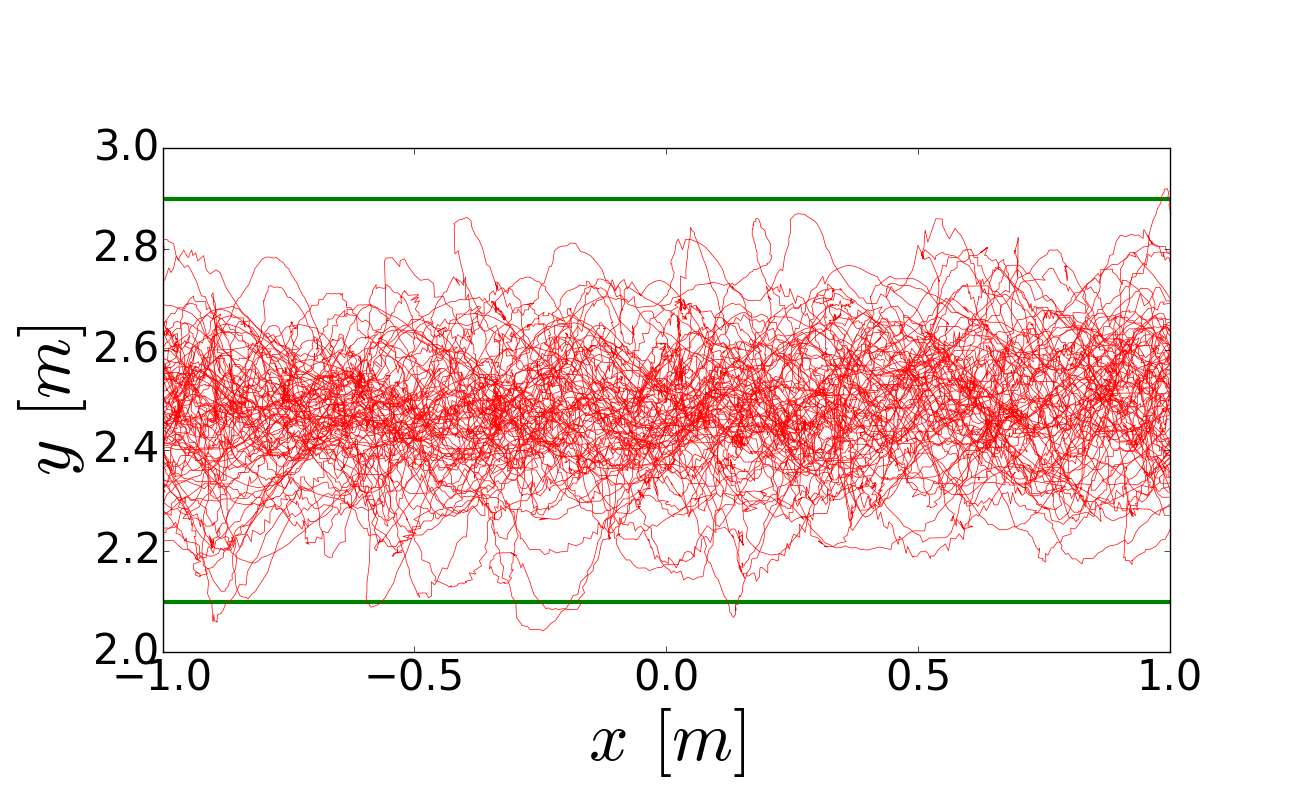}}
 \caption{\label{fig-traj}Pedestrian trajectories from the experiment. The left figure is the trajectory in the whole ring and the right figure is enlargement of the specific region $x \in$ [-1.0 $m$, 1.0 $m$] and $y \in$ [2.1 $m$, 2.9 $m$].}
 \end{figure}

Two cameras were fixed on the fourth floor of a teaching building to record the whole experiment. Since 3D coordinate system is now available in the software $PeTrack$ \cite{Boltes2010, Boltes2013},  videos recorded with a view not perpendicular but slanted to the floor can be used for measuring positions. Three colors of hats were worn by the participants during the experiment to mark their height, which is needed in $PeTrack$ to get more precise 3D position based on image data. The persons who are shorter than 1.6 $m$ wear yellow hat, between 1.6 $m$ and 1.65 $m$ wear blue hat, and red hats are for people taller than 1.65 $m$. FIG. \ref{fig-traj} shows a series of pedestrian trajectories from $PeTrack$ by tracking the hats. To guarantee the accuracy of the data and avoid some mistakes during tracking process, the trajectories are checked and corrected manually after they are extracted automatically from $PeTrack$. It can be seen that the more people in the ring, the more lateral movement happens from the enlargement of the specific region in FIG. \ref{fig-traj}.

\section{Result analysis}\label{analysis}
\subsection{Methods}

We only focus on one dimensional characteristics: longitudinal interactions among pedestrians. Thus in the following analysis 1D movement information is approximated by projecting the 2D coordinates to the central line of the ring. The physical coordinates ($x$, $y$) have been unrolled into a new coordinate denoting the arc-length whose total range is equal to the central circumference ($C$ = 25.7 $m$) of the original ring. With this kind of transformation the lateral oscillations of pedestrian movement are totally neglected. In the new system the maximum of one's coordinate is $x$ = 25.7 $m$.

Microscopically, an individual density can be defined based on headway or Voronoi tessellation. For any pedestrian $i$ at time $t$, the headway $d_{h,i}(t)$ is defined as the distance between the centers of him and his predecessor, while the one dimensional Voronoi distance $d_{V,i}(t)$ is the half distance between the centers of his follower and predecessor. Therefore, the individual density can be calculated based on headway or Voronoi distance:
\begin{equation}\label{equ-rho-headway}
  \rho_{h,i}(t) = 1/d_{h,i}(t)
\end{equation}
\begin{equation}
  \rho_{V,i}(t) = 1/d_{V,i}(t)
\end{equation}
Meanwhile, the individual instantaneous velocity is calculated as:
\begin{equation}\label{equ-v-individual}
  v_i(t) =  \frac{x_i(t+\Delta t/2)-x_i(t-\Delta t/2)}{\Delta t}\emph{}
\end{equation}
where $x_i(t)$ is the coordinate of pedestrian $i$ in the new coordination system at time $t$. $\Delta t$ is the time interval and 0.4 $s$ is used in the paper.

Macroscopically, the average density $\rho(t)$ and velocity $v(t)$ in a measurement length $l_m$ at time $t$ are calculated based on Voronoi method as \cite{Zhang2014}:
\begin{equation}
  \rho(x,t) =  \rho_{V,i}(t)~and~v(x,t)=v_i(t)~if~x\in l_m
\end{equation}

\begin{equation}\label{equ-rho_V}
\rho(t)=\frac{\int{\rho(x,t)dx}}{l_m}
\end{equation}

\begin{equation}\label{equ-v_V}
v(t)=\frac{\int{v(x,t)dx}}{l_m}
\end{equation}

\subsection{Time-space diagram}
\begin{figure*}
 \centering\subfigure[Young-05]{
 \includegraphics[width=0.32\textwidth]{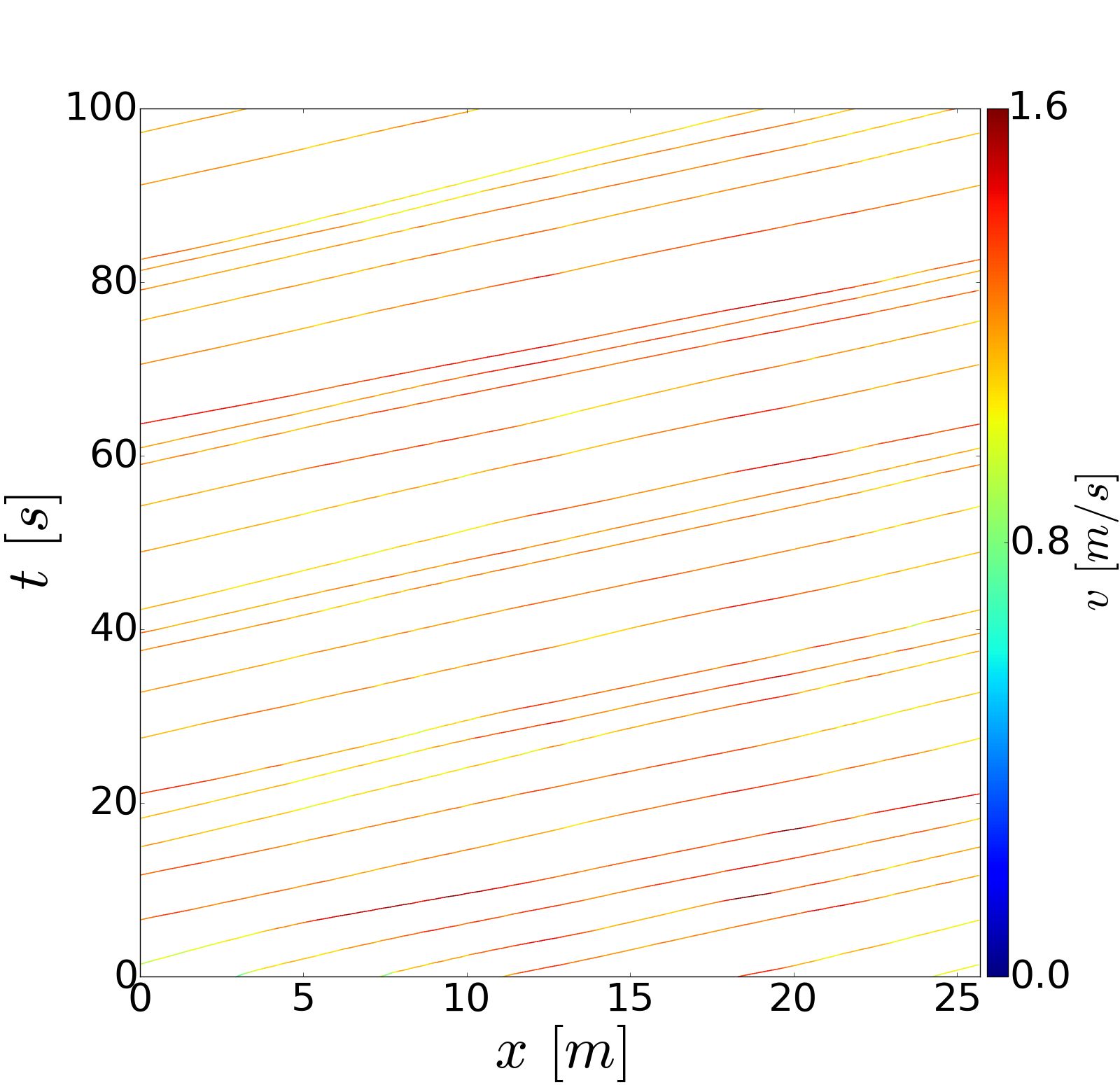}}
 \centering\subfigure[Mixed-06]{
 \includegraphics[width=0.32\textwidth]{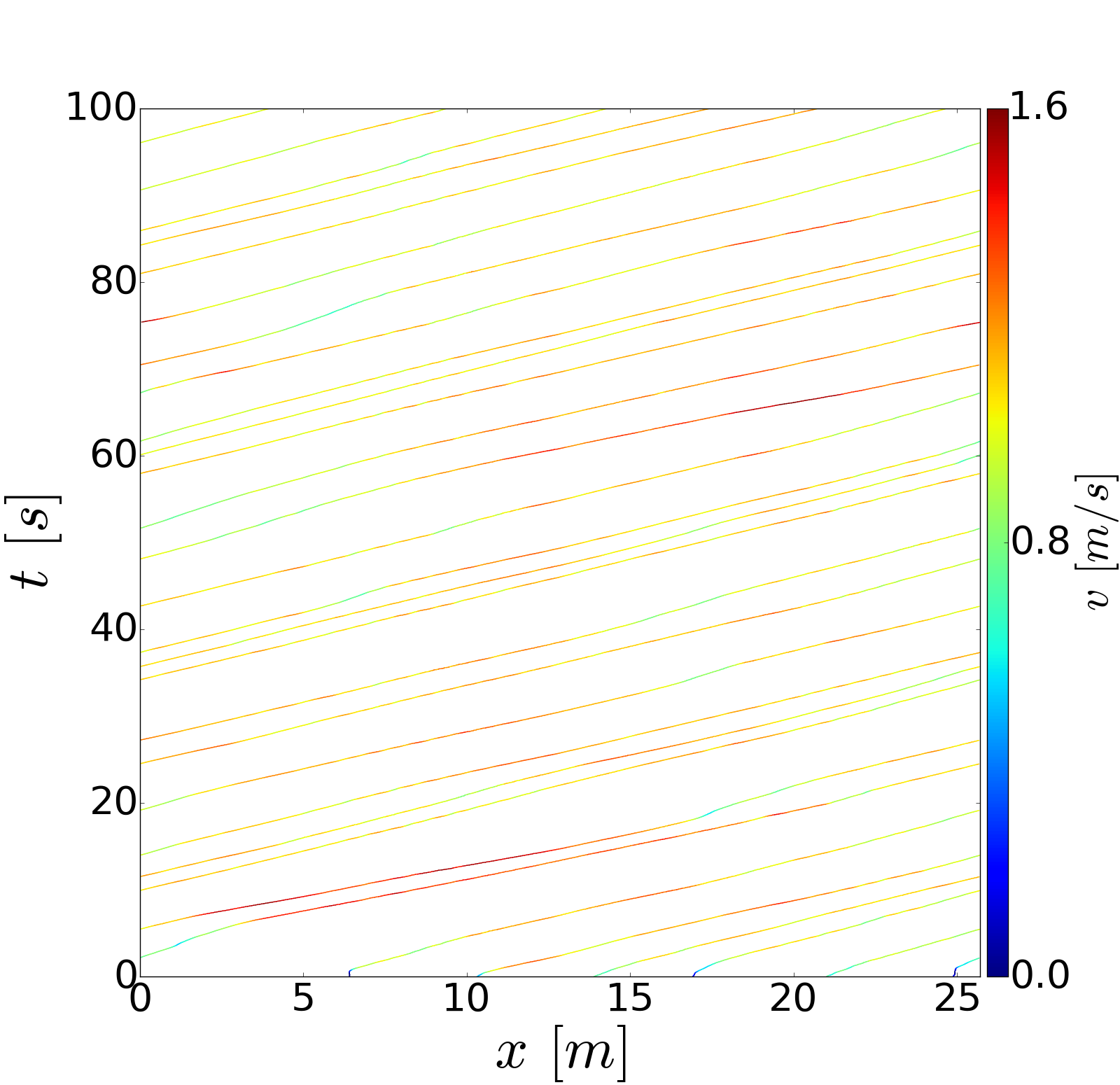}}
 \centering\subfigure[Old-06]{
 \includegraphics[width=0.32\textwidth]{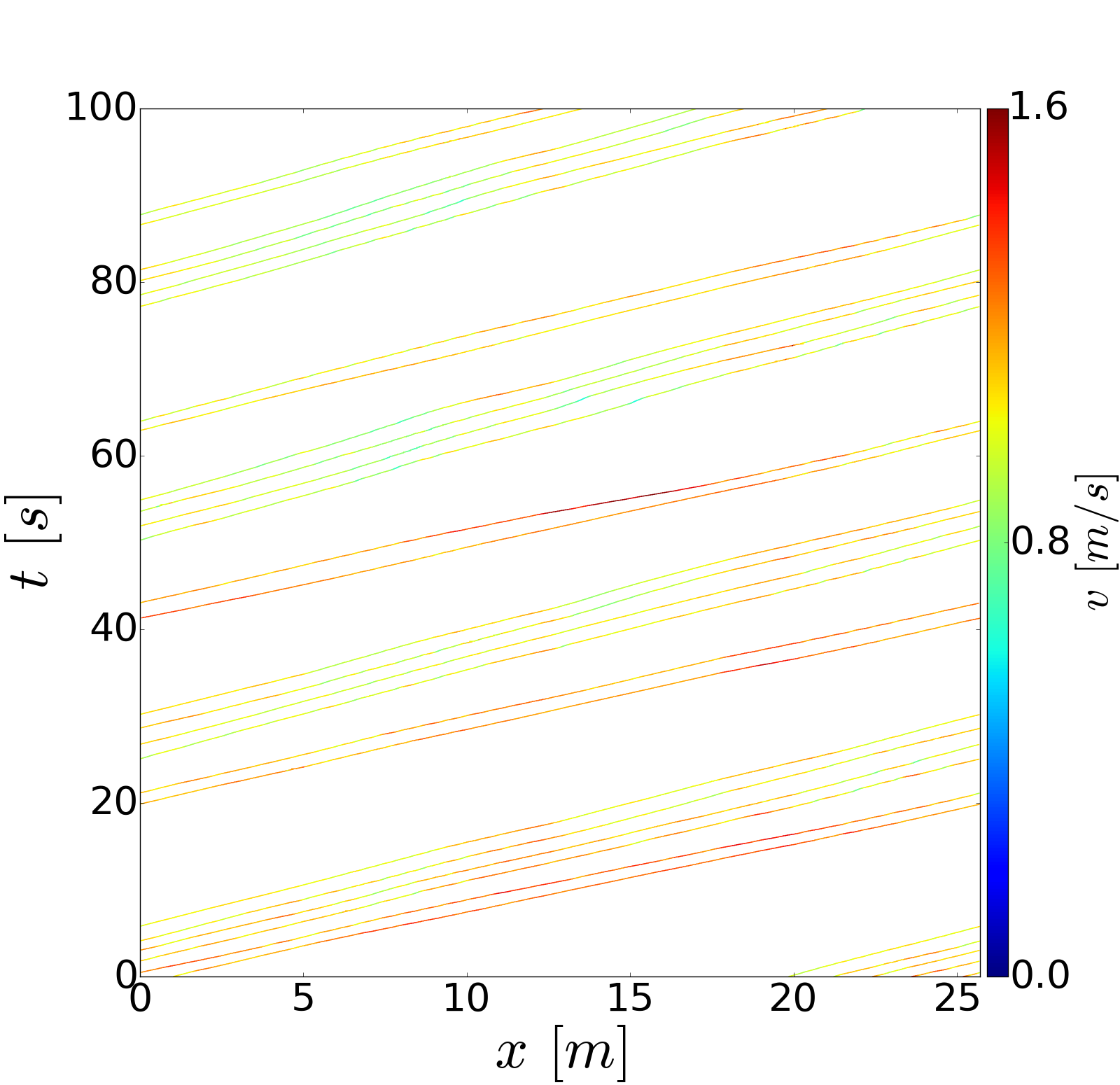}}
 \caption{\label{fig-timespace-low}The time-space diagrams at low density.}
 \end{figure*}

\begin{figure*}
 \centering\subfigure[Young-25]{
 \includegraphics[width=0.32\textwidth]{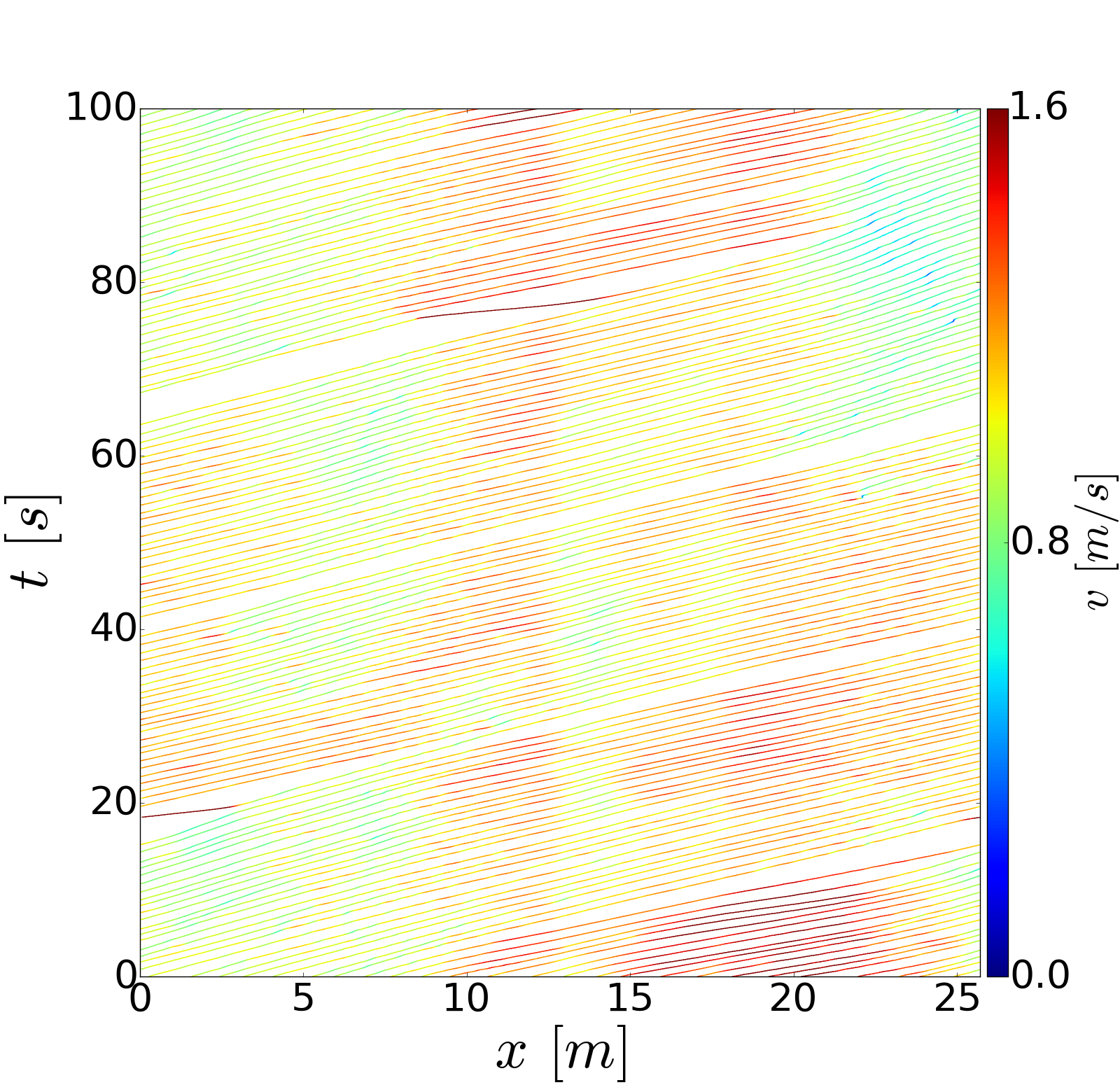}}
 \centering\subfigure[Mixed-26]{
 \includegraphics[width=0.32\textwidth]{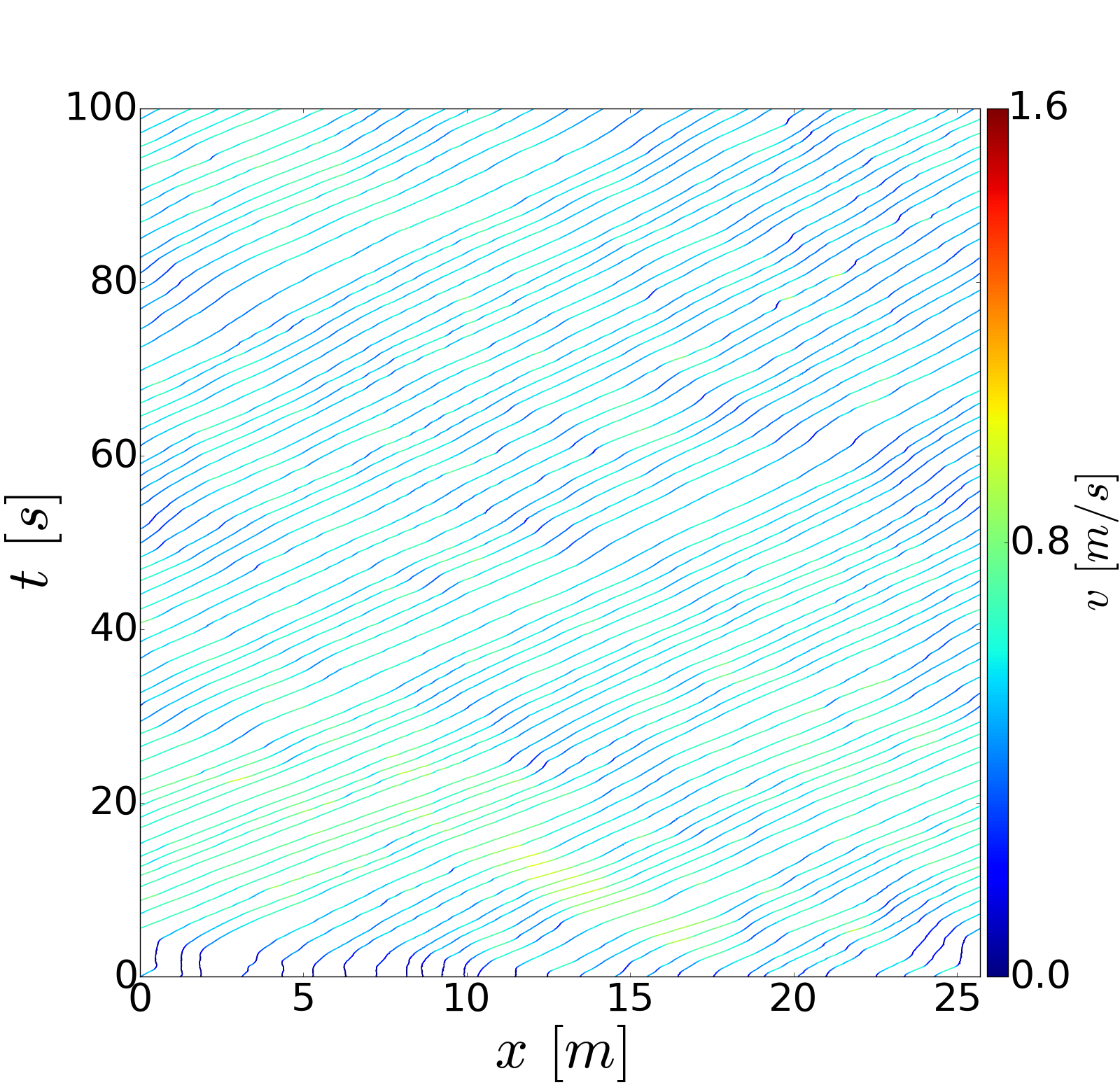}}
 \centering\subfigure[Old-26]{
 \includegraphics[width=0.32\textwidth]{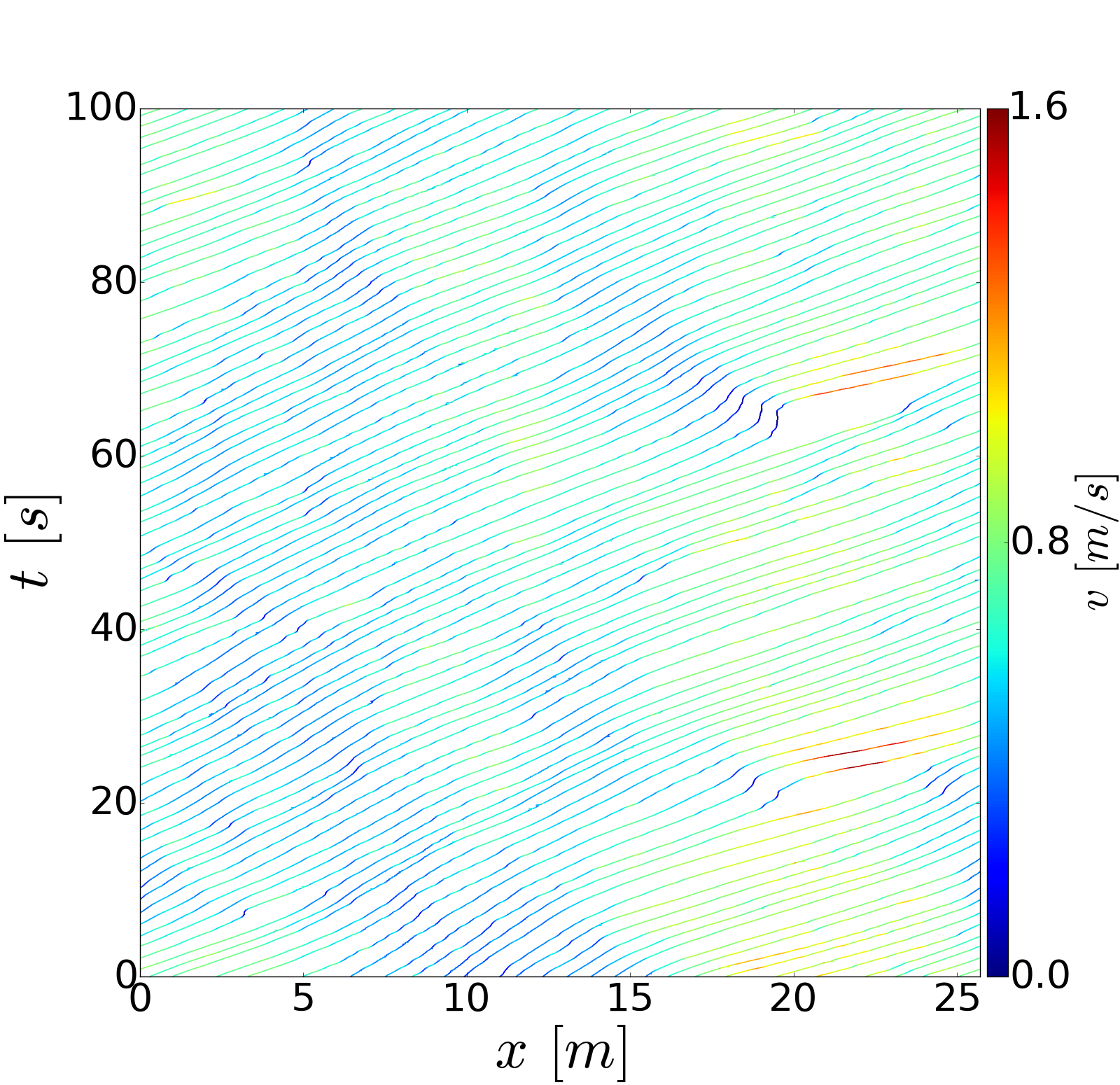}}
 \caption{\label{fig-timespace-medium}The time-space diagrams at medium density.}
 \end{figure*}

FIG.\ref{fig-timespace-low} and FIG.\ref{fig-timespace-medium} show the time-space diagrams for runs at low density ($\rho_g \approx 0.2~m^{-1}$) and medium density ($\rho_g = 1.0 ~m^{-1}$) respectively. The instantaneous velocities $v_i(t)$ are displayed color coded directly in the graph. At low density case, the young student group has a mean velocity of  $1.21 \pm 0.11~m/s$ standard deviation, while it is $1.04 \pm 0.14~m/s$ and $1.07 \pm 0.15~m/s$ for old and mixed group respectively in FIG.\ref{fig-timespace-low}.  At such low density range, pedestrians can move with their free velocities. Due to different movement abilities, the free velocity of young students is higher than that of old pedestrians. The average velocity for the mixed group is a little higher than that of the old pedestrians. At medium density situation, the young student group still has the highest velocity (a mean of $1.05 \pm 0.20~m/s$ standard deviation) among three groups. However, the velocity of mixed group is $0.54 \pm 0.13~m/s$ in FIG.\ref{fig-timespace-medium}(b) and $0.62 \pm 0.15~m/s$ for the old group in FIG.\ref{fig-timespace-medium}(c). The mixed group has a lower velocity than the old group (T-test, P-value $< 2.2\times10^{-16}$), which indicates that the students' movement in the mixed crowd is not simply determined by the old pedestrians (slow object) at medium density situation. Conversely, the complex interaction between the young students and old people makes them move slower.

\begin{figure}
 \centering\subfigure[Young-45]{
 \includegraphics[width=0.23\textwidth]{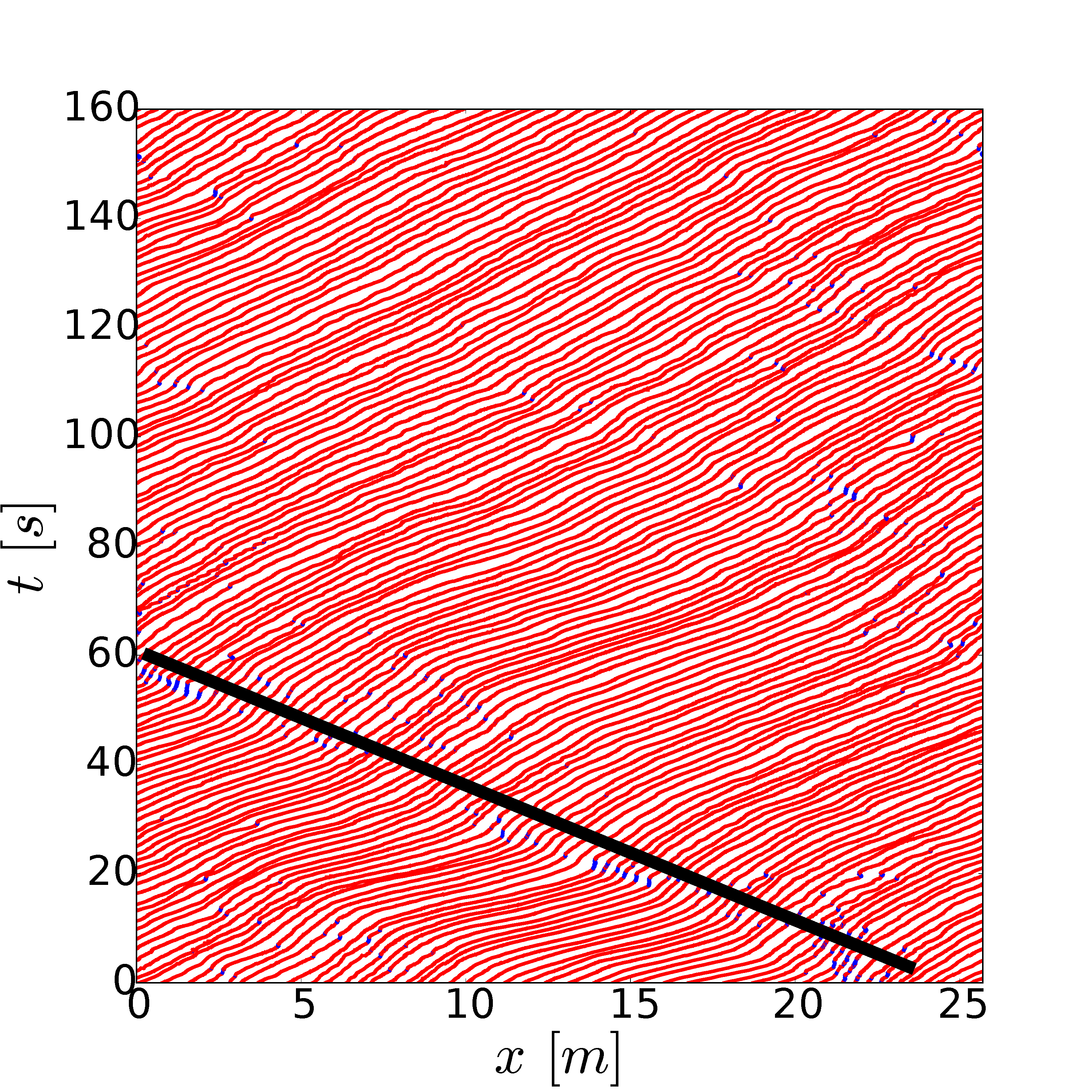}}
 \centering\subfigure[Mixed-46]{
 \includegraphics[width=0.23\textwidth]{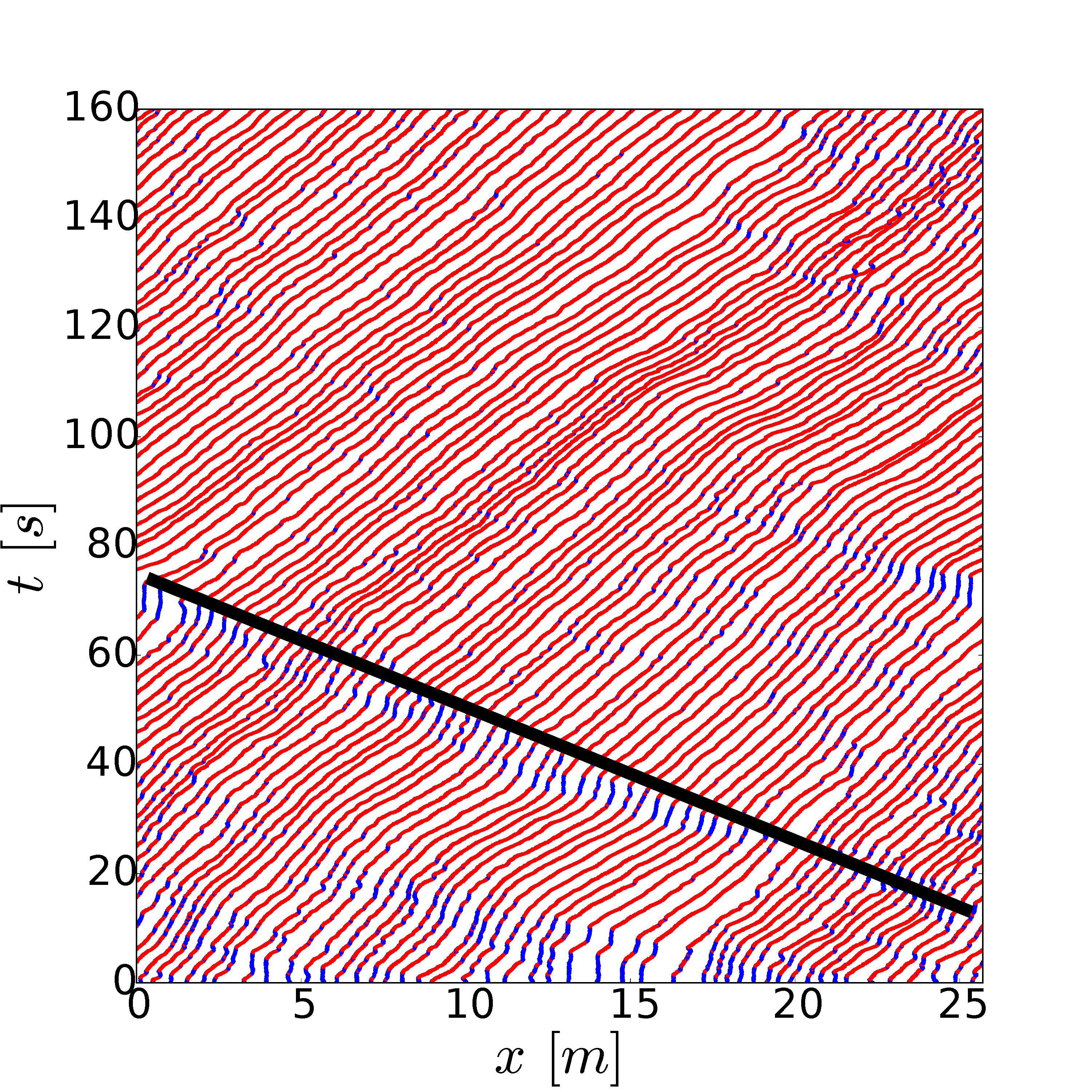}}
 \centering\subfigure[Young-61]{
 \includegraphics[width=0.23\textwidth]{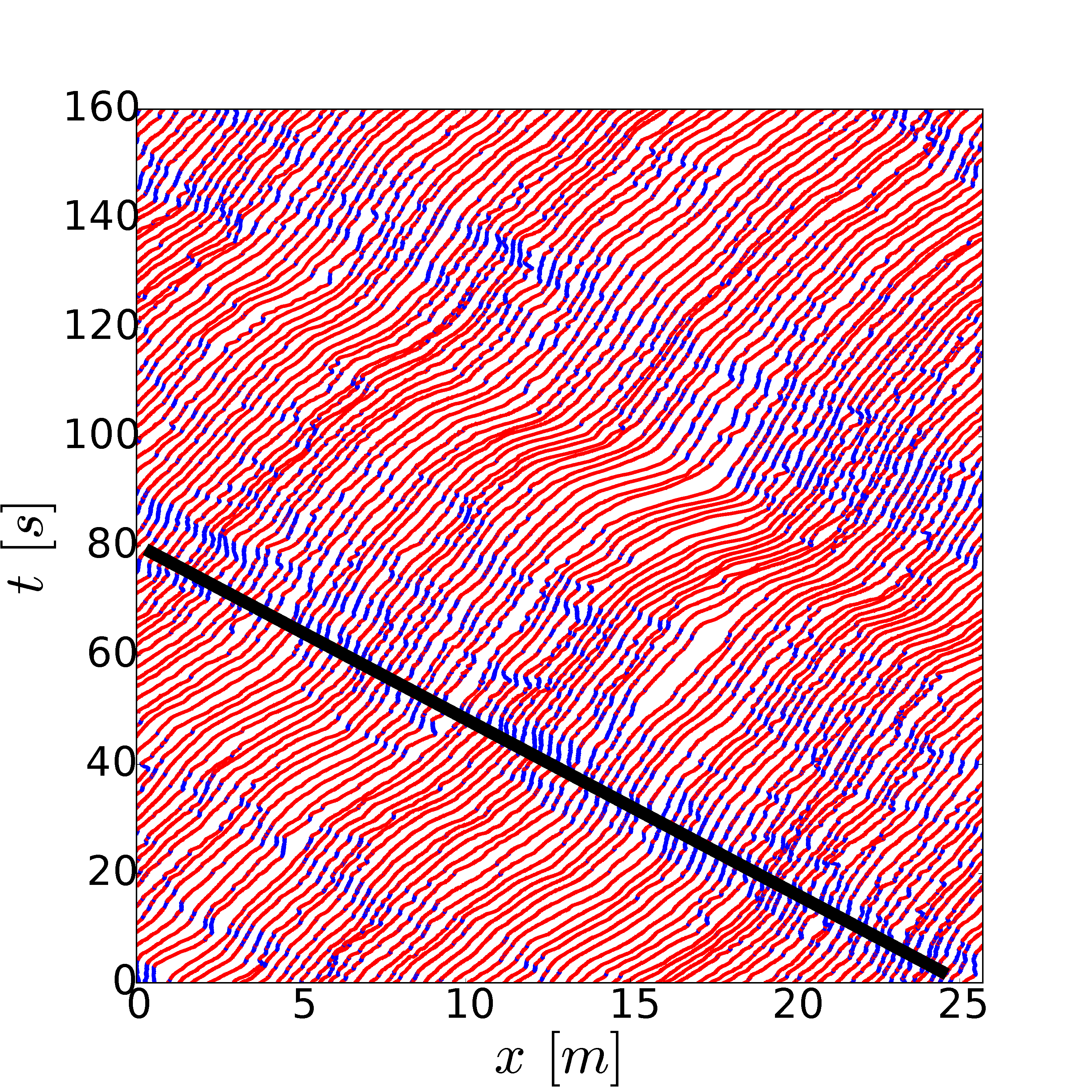}}
  \centering\subfigure[Mixed-60]{
 \includegraphics[width=0.23\textwidth]{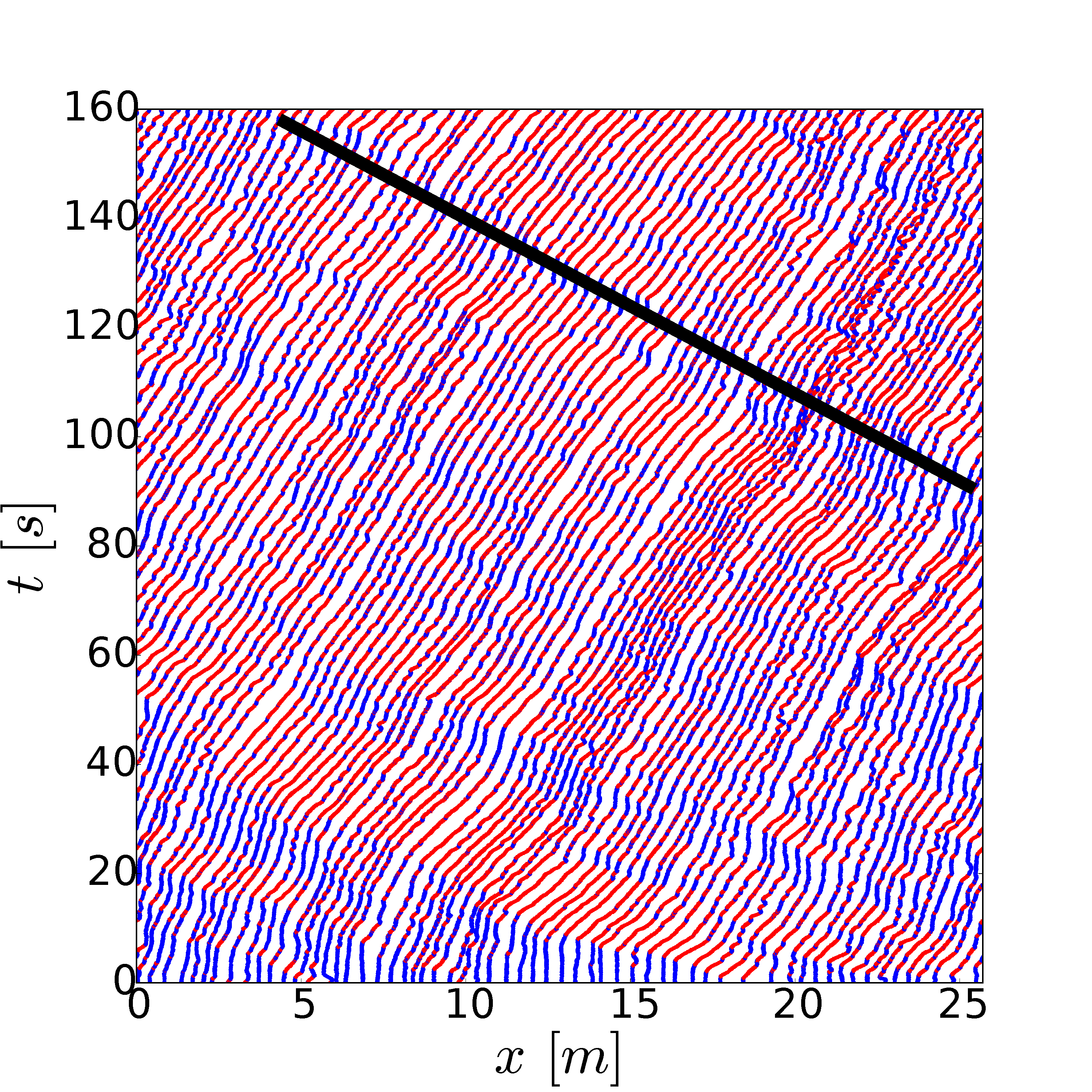}}
 \centering\subfigure[Young-66]{
 \includegraphics[width=0.23\textwidth]{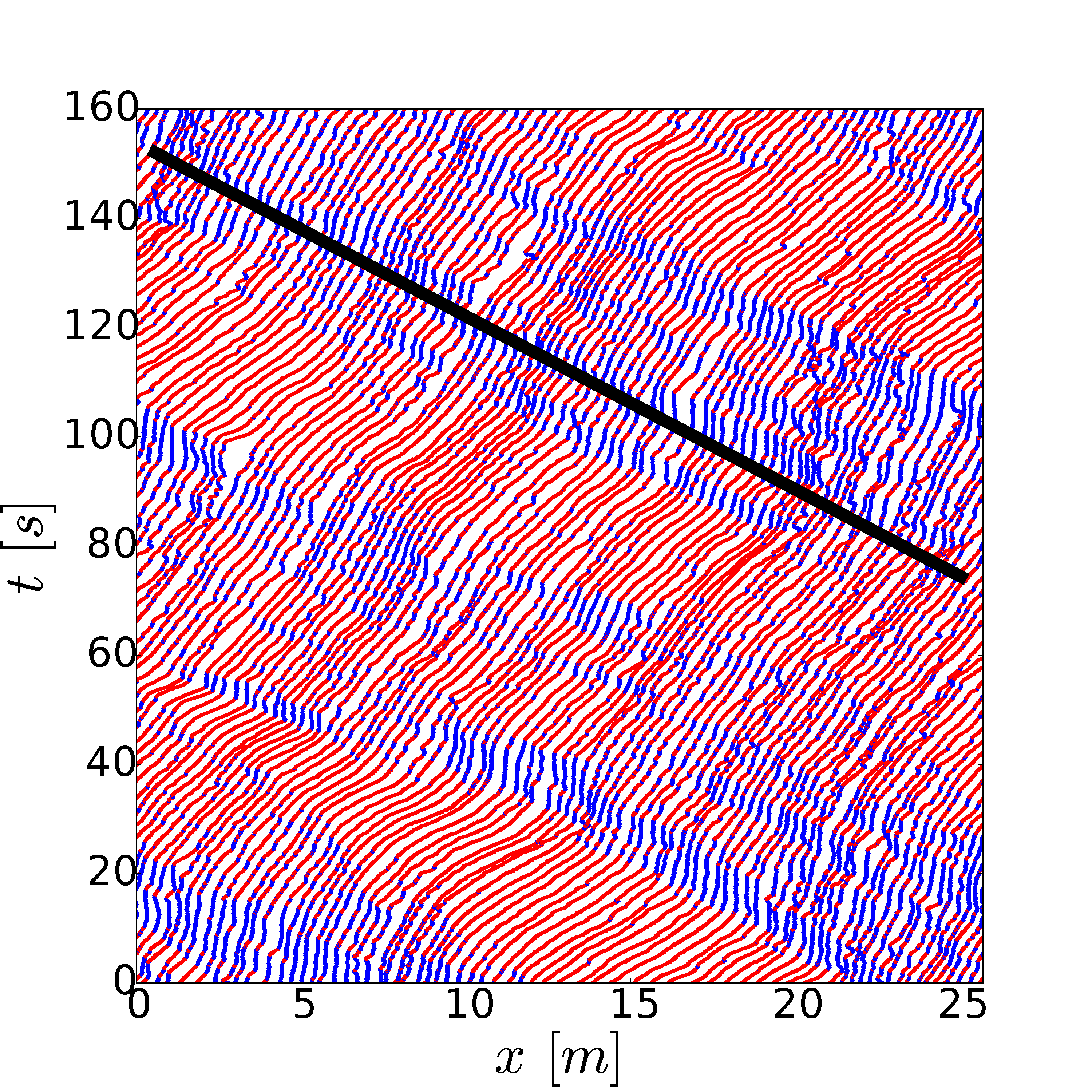}}
 \centering\subfigure[Young-71]{
 \includegraphics[width=0.23\textwidth]{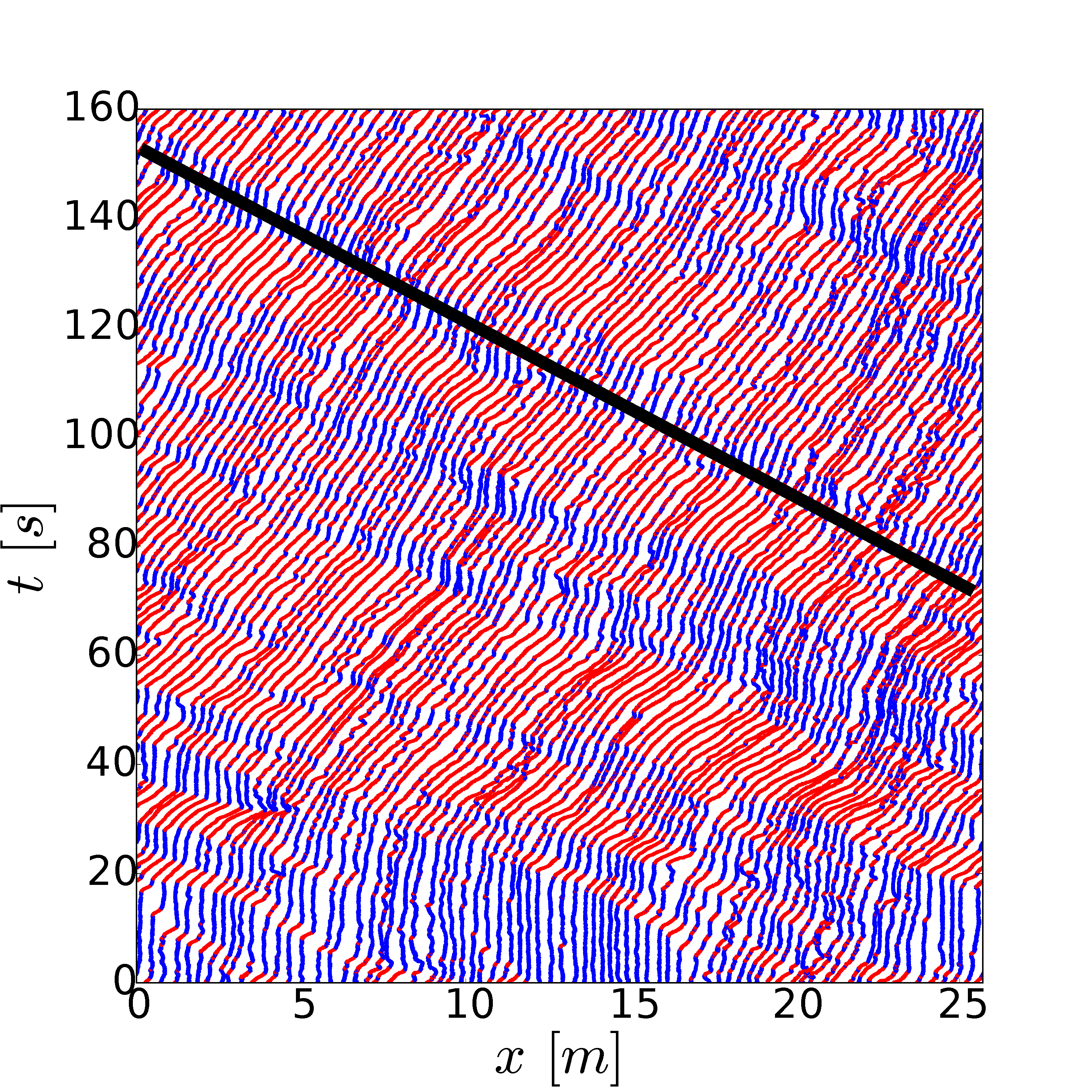}}
 \caption{\label{fig-timespace-high}The stop-and-go waves under high density. The individual velocity is shown in red for $v_i(t) > 0.1~m/s$ and blue for $v_i(t) < 0.1~m/s$ under which pedestrians are regarded as in stop phase. The absolute value of the slope of the black line represents the propagation velocity $v_{jam}$ of the stop-and-go wave. For (a) and (b), the global density is about $1.75~m^{-1}$ and $v_{jam}$ is about 0.4 $m/s$. For the runs (c), (d), (e) and (f), the global density is higher than 2.3 $m^{-1}$ and $v_{jam}$ are all about 0.3 $m/s$.}
 \end{figure}

 Under high density situations for $\rho_g > 1.75~m^{-1}$, pedestrians' velocities are not homogeneous anymore over space and time and the stop-and-go waves are observed. Unfortunately, we only have experimental data for mixed group and young student group at this density range. To visualize the jams or the potential stop-and-go, the time-space diagram is displayed in another way. As shown in FIG.\ref{fig-timespace-high}, the individual velocity is shown in red for $v_i(t) > 0.1~m/s$ and blue for other values. When $v_i(t) < 0.1~m/s$ pedestrians can be regarded as in stop phase, which is nothing but jam. Firstly, we compare the emergence of the jam in different kind of groups. From FIG.\ref{fig-timespace-high}(a) to (d), it shows that more jams occur in the mixed group than that of student group at the same global density situation. For both groups, the sizes of jams increase with the increment of density. In addition, it is found that the stop-and-go wave propagates in the opposite direction to the movement of pedestrians with certain velocity $v_{jam}$, which is indicated by the absolute value of slope of the black line in the diagram. For the runs 'Young-45' and 'Mixed-46' ($\rho_g = 1.75~m^{-1}$ and 1.78 $m^{-1}$), $v_{jam}$ is about 0.4 $m/s$.  However, it is roughly 0.3 $m/s$ for all the other runs whose global density $\rho_g > 2.3~m^{-1}$. The propagation velocity of stop and go wave is related to the relative speed between moving and non-moving pedestrians, and it is larger in slightly crowed situation (0.28 $m/s$ for 'Young-45' and 0.26 $m/s$ for 'Mixed-46') than that in severely congested situation (0.13 $m/s$ for 'Young-66' and 0.13 $m/s$ for 'Young-71'). This maybe the reason why different $v_{jam}$ are obtained under different density ranges.

\subsection{Fundamental diagram}

\begin{figure}
 \includegraphics[width=0.40\textwidth]{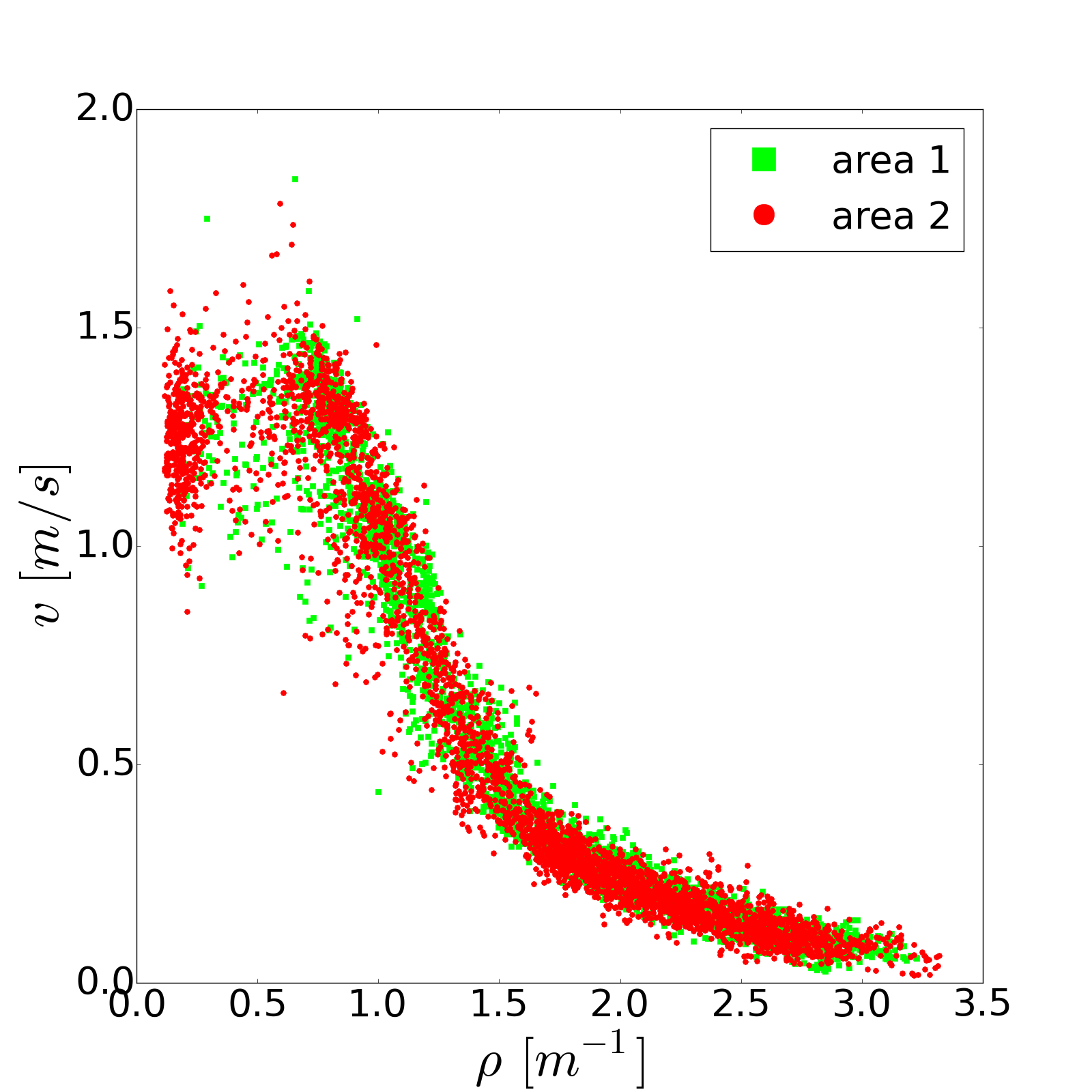}
 \caption{\label{fig-FD-measurearea}Comparison of the fundamental diagram for the student group at different measurement areas. Area 1 means the area [7 $m$, 11 $m$] and area 2 is [13.85 $m$, 17.85 $m$].}
 \end{figure}

In this section we focus on the fundamental diagrams for the movement of different groups. Note that in our analysis only data under steady state are used. The data at the beginning and the end of each run are removed to avoid the effect of initial condition on the results. Firstly, the influence of different locations of measurement area is studied by using the data from young student group. Two measurement areas with the same size $x \in$ [7 $m$, 11 $m$] and [13.85 $m$, 17.85 $m$] are selected, which corresponds to one locating on the curve and the other one on the straight line. The mean density and velocity over the areas are calculated frame by frame. But only one sample per second is used in comparison to guarantee the independency of the samples. As displayed in FIG.\ref{fig-FD-measurearea}, no difference is observed in these two areas. It seems the macroscopic properties for one dimensional movement are the same no matter pedestrians move along straight line or around a circle.

\begin{figure*}
 \centering\subfigure[Density-velocity]{
 \includegraphics[width=0.4\textwidth]{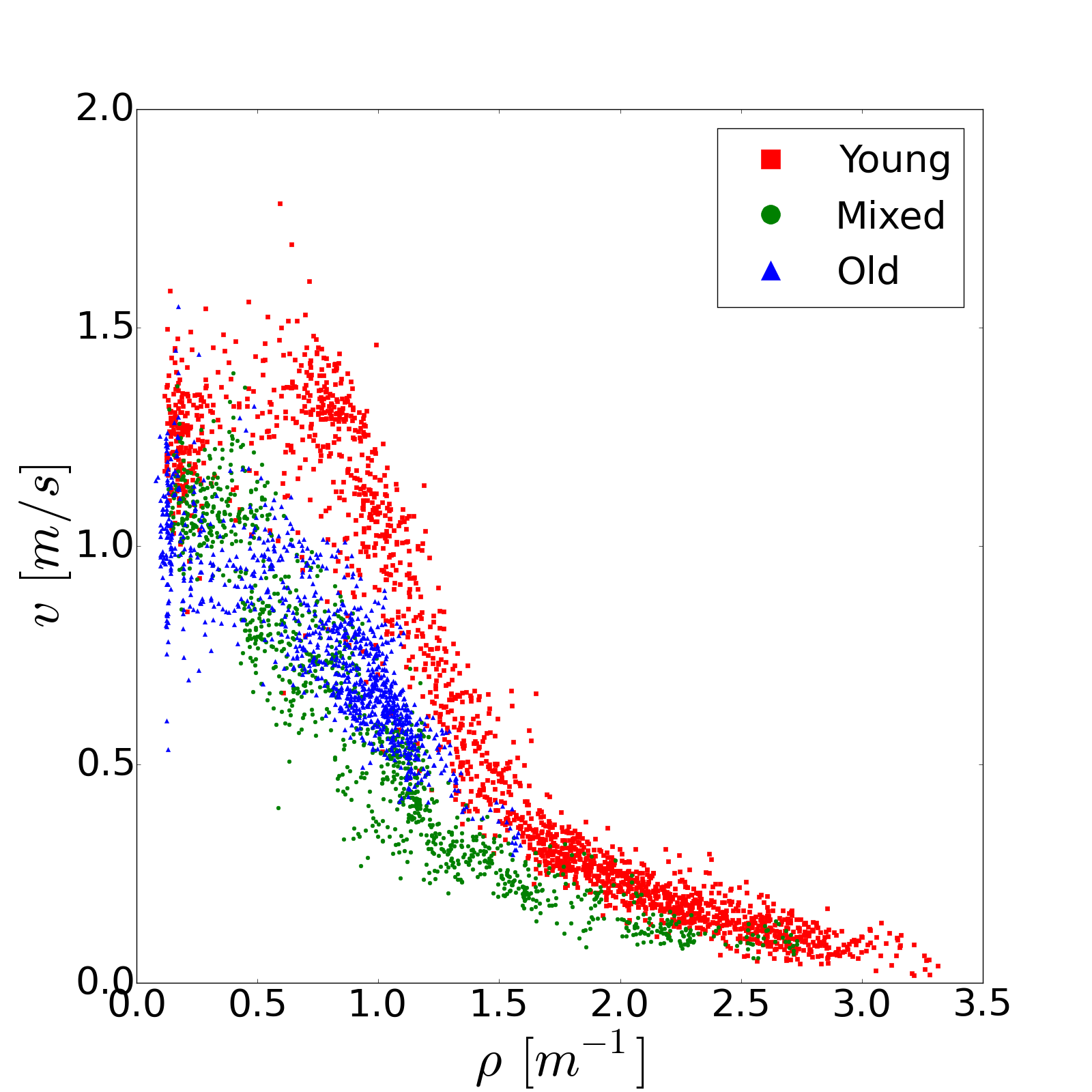}}
 \centering\subfigure[Density-flow]{
 \includegraphics[width=0.4\textwidth]{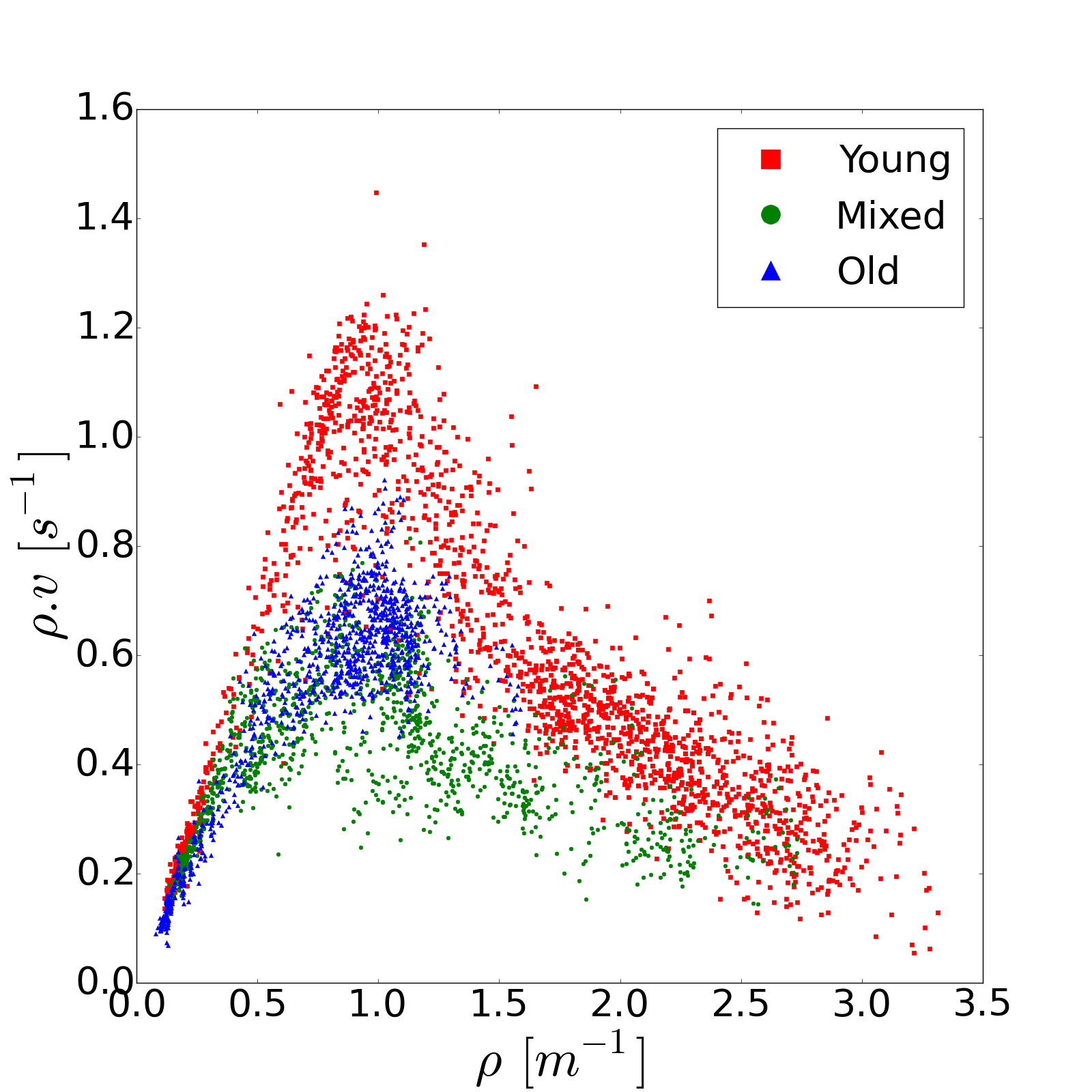}}
 \caption{\label{fig-FD-groups}Comparison of the fundamental diagrams obtained from Voronoi methods (equation (\ref{equ-rho_V}) and (\ref{equ-v_V})). The red, blue and green points stand young student, old and mixed group respectively.}
 \end{figure*}

 Based on the above comparison, we use $l_m = 4~m$ from 13.85 $m$ to 17.85 $m$ as the measurement area for all these three groups in the following analysis. Due to the stronger movement ability, higher activeness and flexibility, young students move faster than the old under the observed density ranges (see FIG.\ref{fig-FD-groups}). At the same density situation, the velocity of young student group is always higher than that of mixed group. Because of the limited test persons, only data for density $\rho < 1.2~m^{-1}$ are obtained from the experiment of old group. Consistent with previous analysis, the velocity of mixed group is a little lower than that of old group for the density between 0.5 and 1.2 $m^{-1}$, whereas it is higher for $\rho < 0.5~m^{-1}$. Besides,  the flow increases monotonically with the increment of density for all groups but reaches different peaks (1.3 $s^{-1}$, 0.9 $s^{-1}$ and 0.7 $s^{-1}$ for young, old and mixed group respectively) around $0.9~m^{-1}$. When density $\rho$ is higher than $0.9~m^{-1}$, the streams go into congested state and the flow decreases. However, the decline rates of the flow with the increasing density become different when the density is higher than $1.8~m^{-1}$. For $\rho>1.8~m^{-1}$, stop-and-go waves occur frequently and dominate pedestrian motion. The decline rate which can reflect the propagation velocity of the stop-and-go wave becomes smaller under this condition. This is accordant with the findings obtained from the time-space diagram.

  \begin{figure*}
 \centering\subfigure[Young student group]{
 \includegraphics[width=0.32\textwidth]{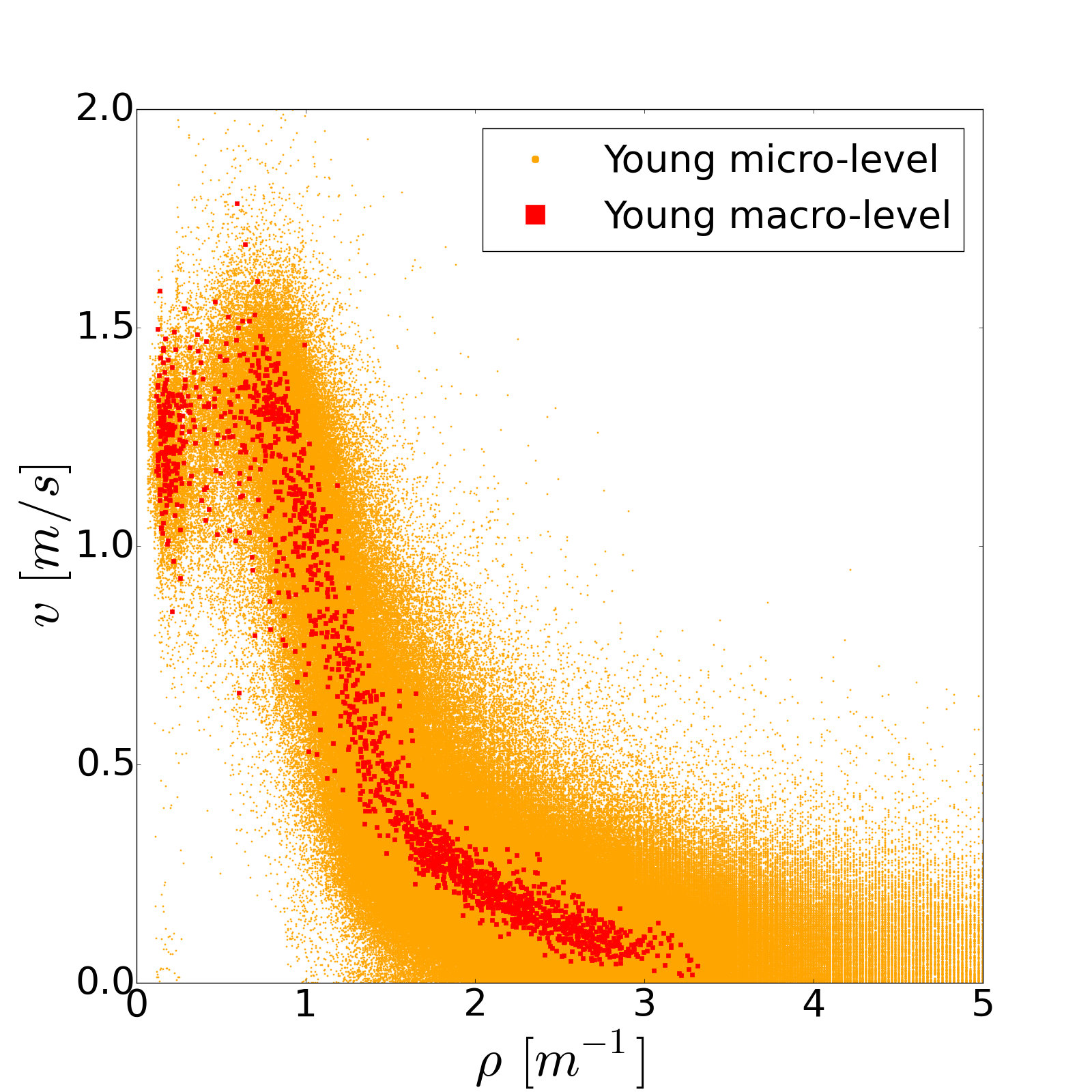}}
 \centering\subfigure[Mixed group]{
 \includegraphics[width=0.32\textwidth]{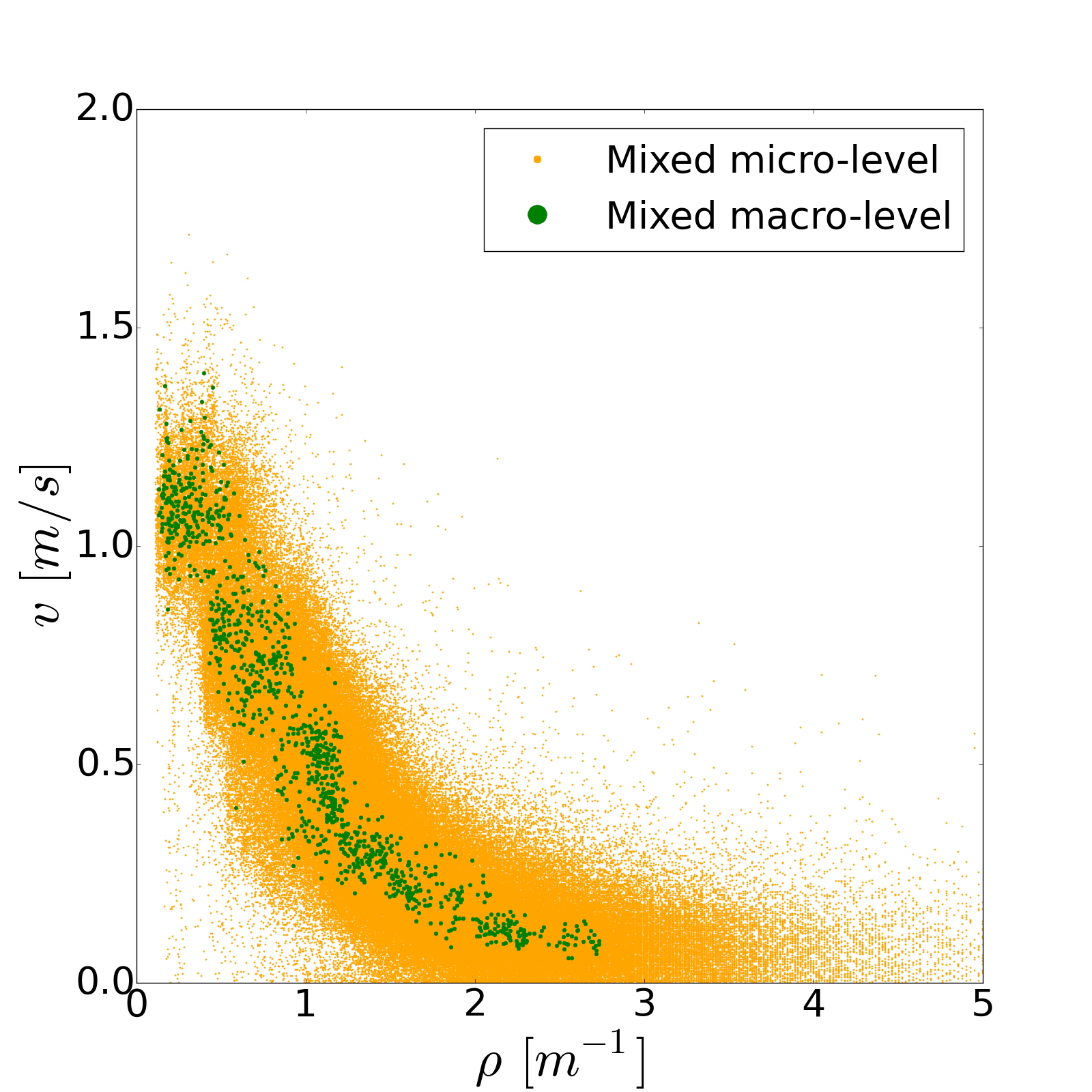}}
 \centering\subfigure[Old group]{
 \includegraphics[width=0.32\textwidth]{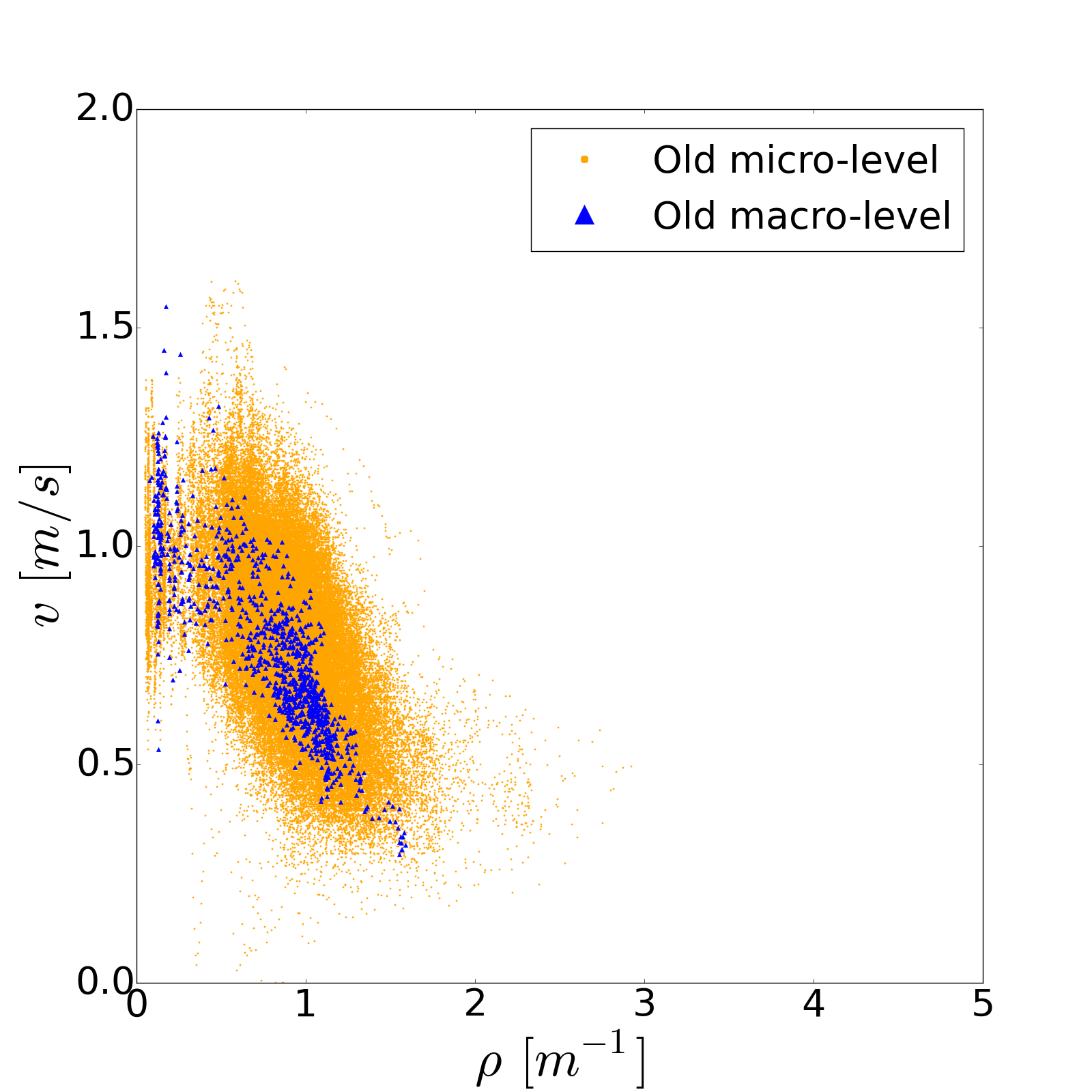}}
 \caption{\label{fig-FD-micro-macro}Comparison of the fundamental diagrams on micro and macro levels for different groups. Note that on a micro-level, the density is obtained based on headway $d_{h,i}(t)$  and velocity is the instantaneous velocity of individual as calculated in equation (\ref{equ-rho-headway}) and (\ref{equ-v-individual}). On a macro-level, the density and velocity are calculated with equation (\ref{equ-rho_V}) and (\ref{equ-v_V}).}
 \end{figure*}

FIG.\ref{fig-FD-micro-macro} compares the fundamental diagrams obtained from micro and macro ways. At a micro level, the relation between instantaneous individual velocity and the corresponding density which is defined as the inverse of the individual headway is depicted. The fundamental diagrams obtained from these two aspects show similar properties of pedestrian movement but different resolution. Especially for the young student group and mixed group, the maximal density is about 3 $m^{-1}$ at the macro level, whereas several data reach to 5 $m^{-1}$ at the micro level. This is easy to be understood. The pedestrian's trajectory in our work is obtained by tracking the position of pedestrian's head. At high density situation, stop-and-go appears and oscillation of pedestrians' body especially head becomes more frequent, which causes the small headway (high density) for individual pedestrian.

\begin{figure*}
 \centering\subfigure[Density-velocity]{
 \includegraphics[width=0.4\textwidth]{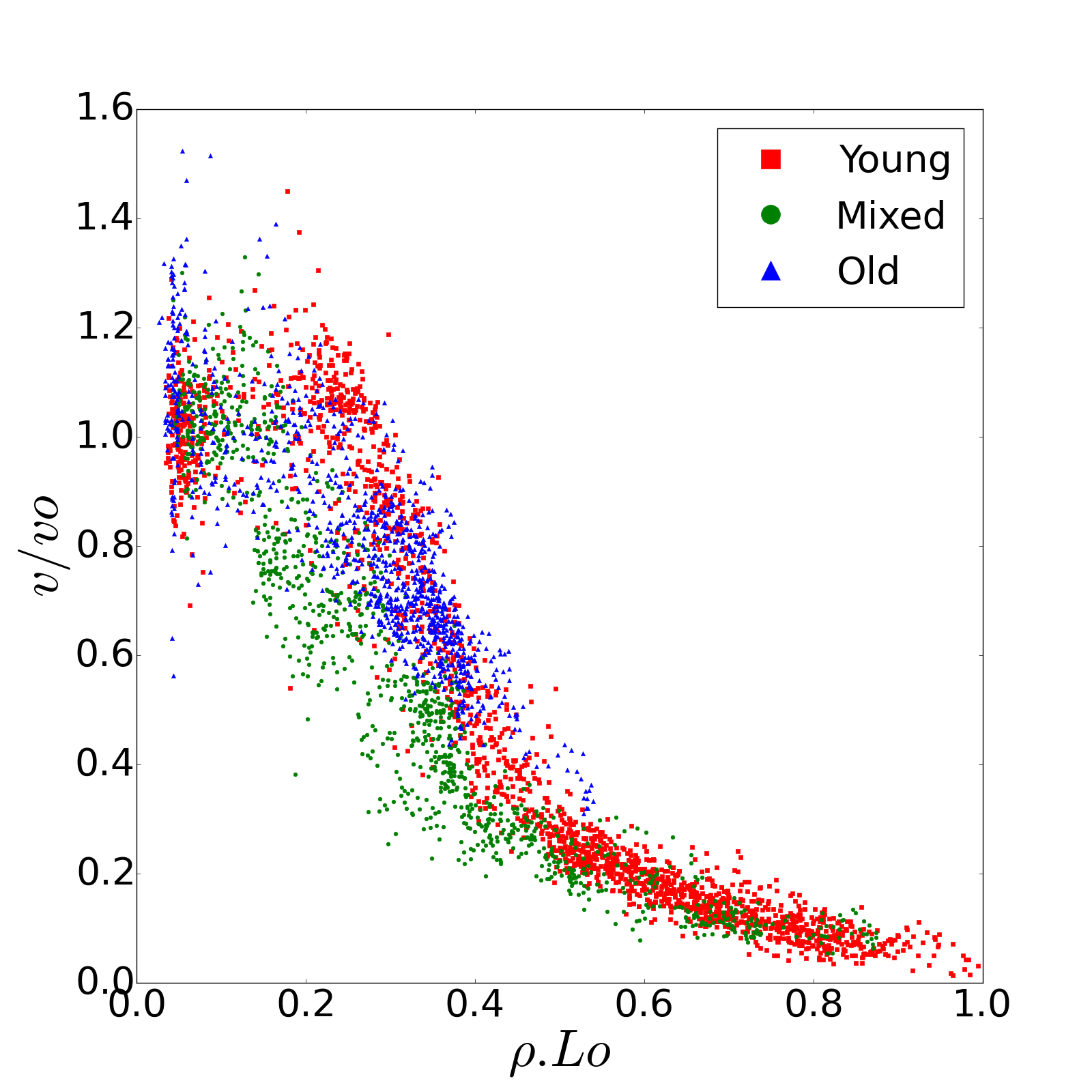}}
 \centering\subfigure[Density-flow]{
 \includegraphics[width=0.4\textwidth]{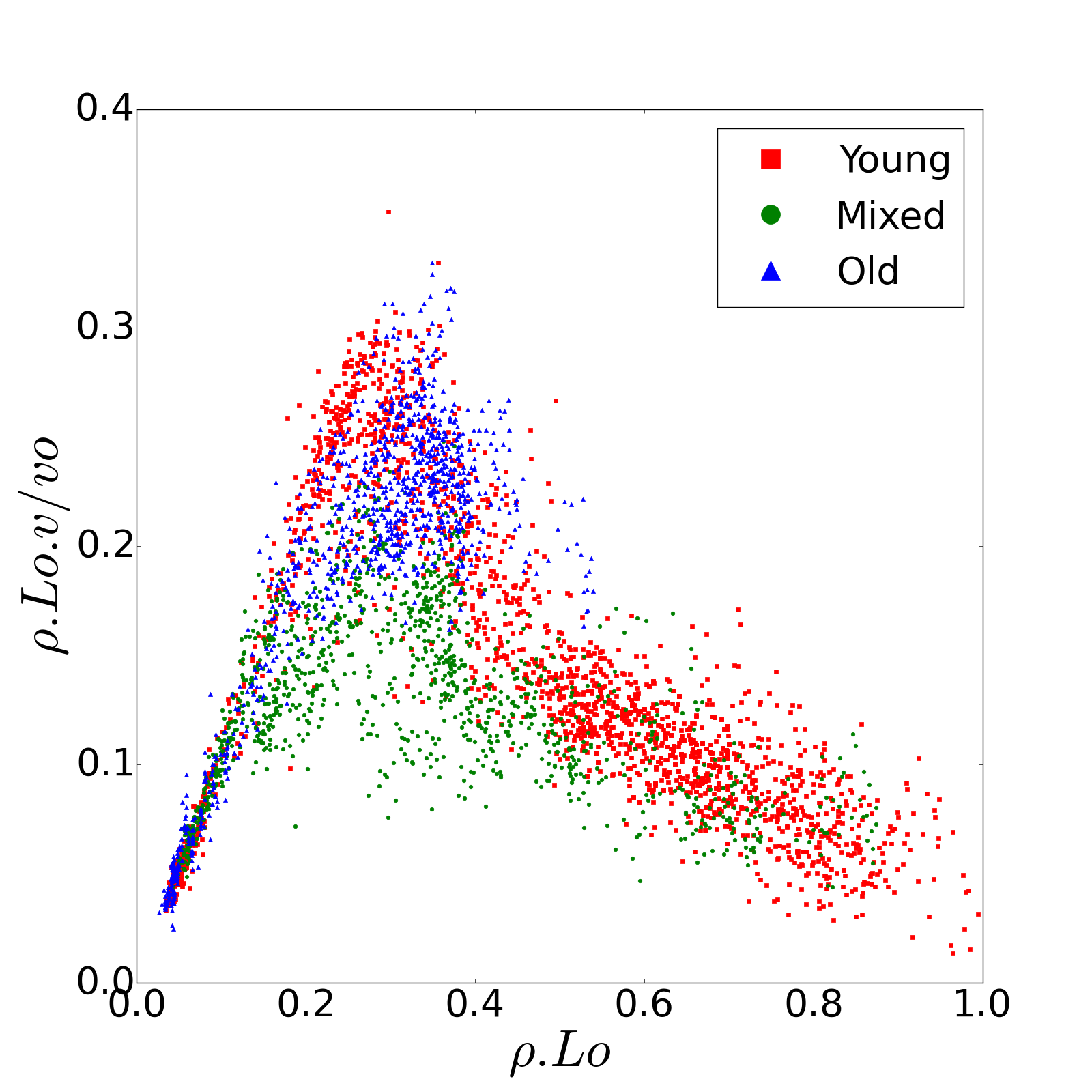}}
 \caption{\label{fig-FD-nondimensional}Comparison of the nondimensionalized fundamental diagram by using the free velocity (1.23 $m/s$, 0.95 $m/s$ and 1.05 $m/s$ for young student, old and mixed group respectively) and body size (0.30 $m$, 0.34 $m$ and 0.32 $m $ for young student, old and mixed group respectively).}
 \end{figure*}

As mentioned in Ref. \cite{Zhang2014}, a uniform and non-dimensional fundamental diagram for pedestrian or vehicle system may exist. Here the same dimensionless method is adopted to make the variables dimensionless and compare the fundamental diagrams of different groups. $v / v_0$ and $\rho \cdot L_0$ are used instead of $v$ and $\rho$ in the nondimensionalized fundamental diagram, where $v_0$ represents the free velocity of pedestrians in the experiment. $v_0$ is set as 1.23 $m/s$, 0.95 $m/s$ and 1.05 $m/s$ for young student group, old group and mixed group respectively. These values are the mean individual velocity of pedestrians under conditions when pedestrian velocity is independent on the corresponding headway (see FIG.\ref{fig-headway-velocity-regimes}). $L_0$ reflects the body size of pedestrians. Before experiment, the chest circumferences of all the test persons are measured. For simplicity in the calculation and analysis, we assume that the shape of pedestrian body's projection on the ground is a circle and use the diameter of the circle as $L_0$. Based on such assumption, we get the average body size $L_0 = 0.30~m$ and $L_0 = 0.34~ m$ for young students and old people. For the mixed group, we use the mean value $L_0 = 0.32~ m$ of the former two sizes. With this nondimensional method, the difference of the three fundamental diagrams seems smaller in both density-velocity and density-flow relations in FIG.\ref{fig-FD-nondimensional}. They agree well with each other when $\rho \cdot L_0 < 0.15$ (absolute free flow state) and $\rho \cdot L_0 > 0.45$ (congested states with stop-and-go waves). However, obvious differences can be observed under medium density situations especially for the mixed group. It seems that the movement properties of the mixed groups are quite complex and are not able to be obtained from other groups by simply scaling with free velocity and body dimension. Regarding to the difference between young student and old people, more empirical data of old group are needed.

 \subsection{Headway-velocity}

\begin{figure*}
 \centering\subfigure[Young student group]{
 \includegraphics[width=0.32\textwidth]{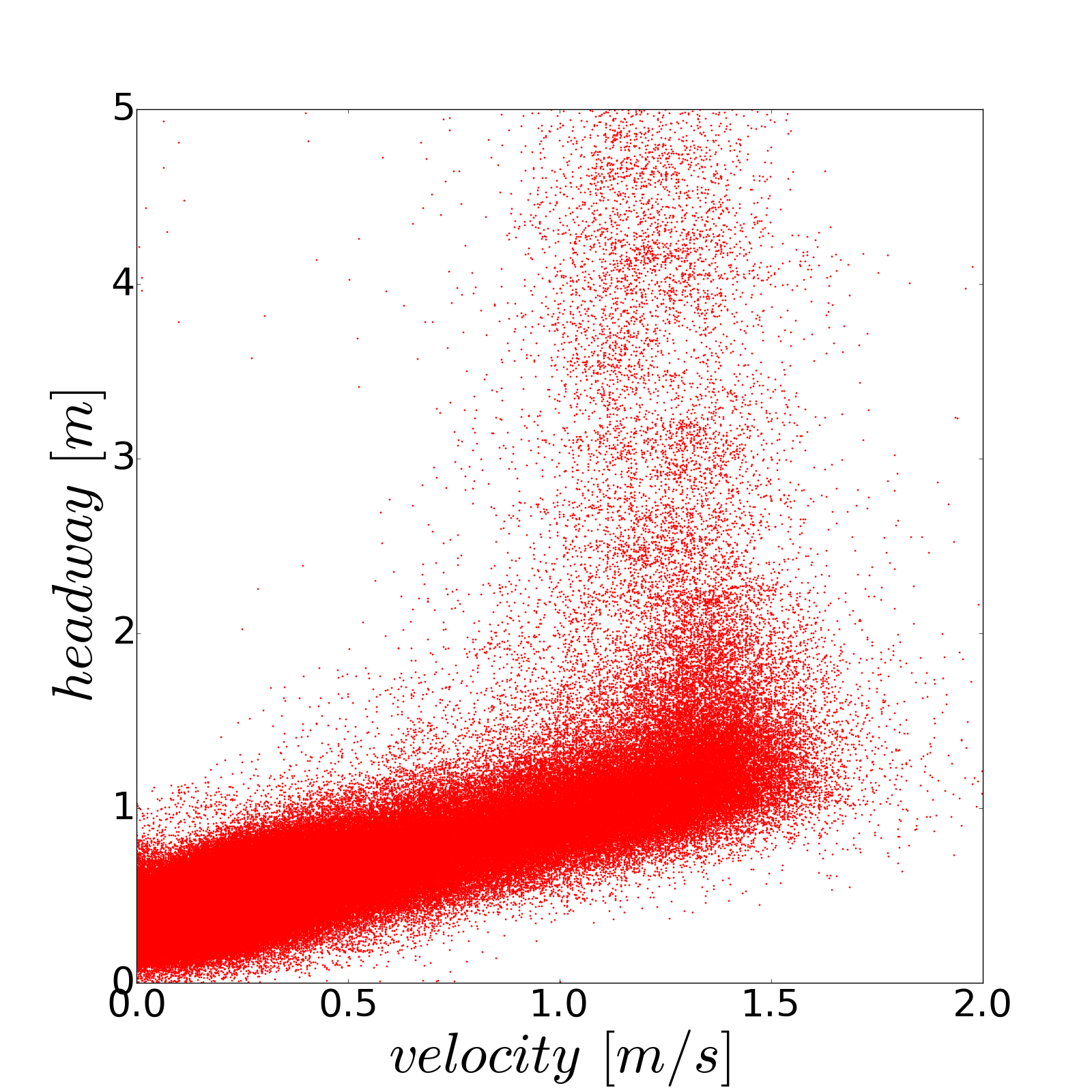}}
 \centering\subfigure[Mixed group]{
 \includegraphics[width=0.32\textwidth]{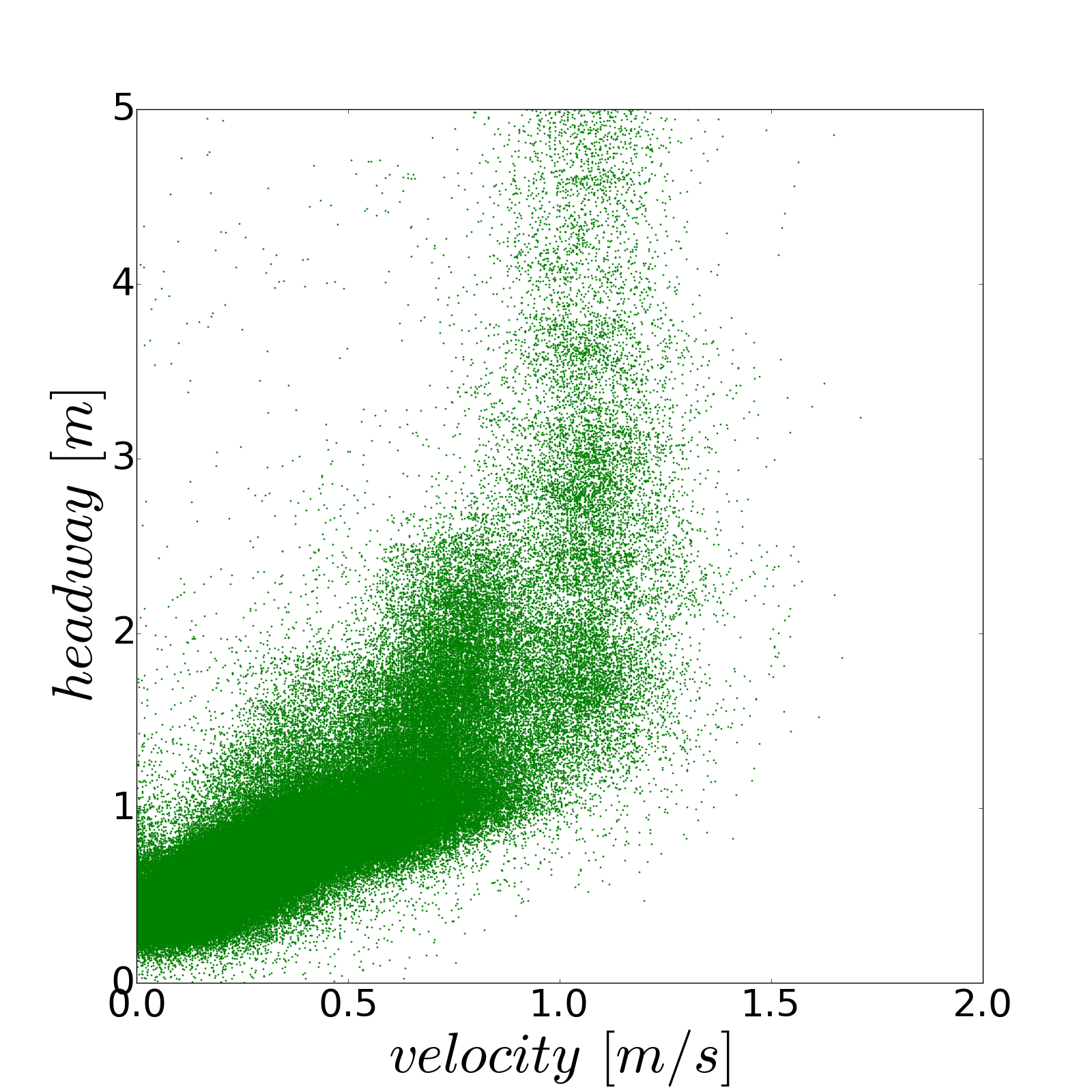}}
 \centering\subfigure[Old group]{
 \includegraphics[width=0.32\textwidth]{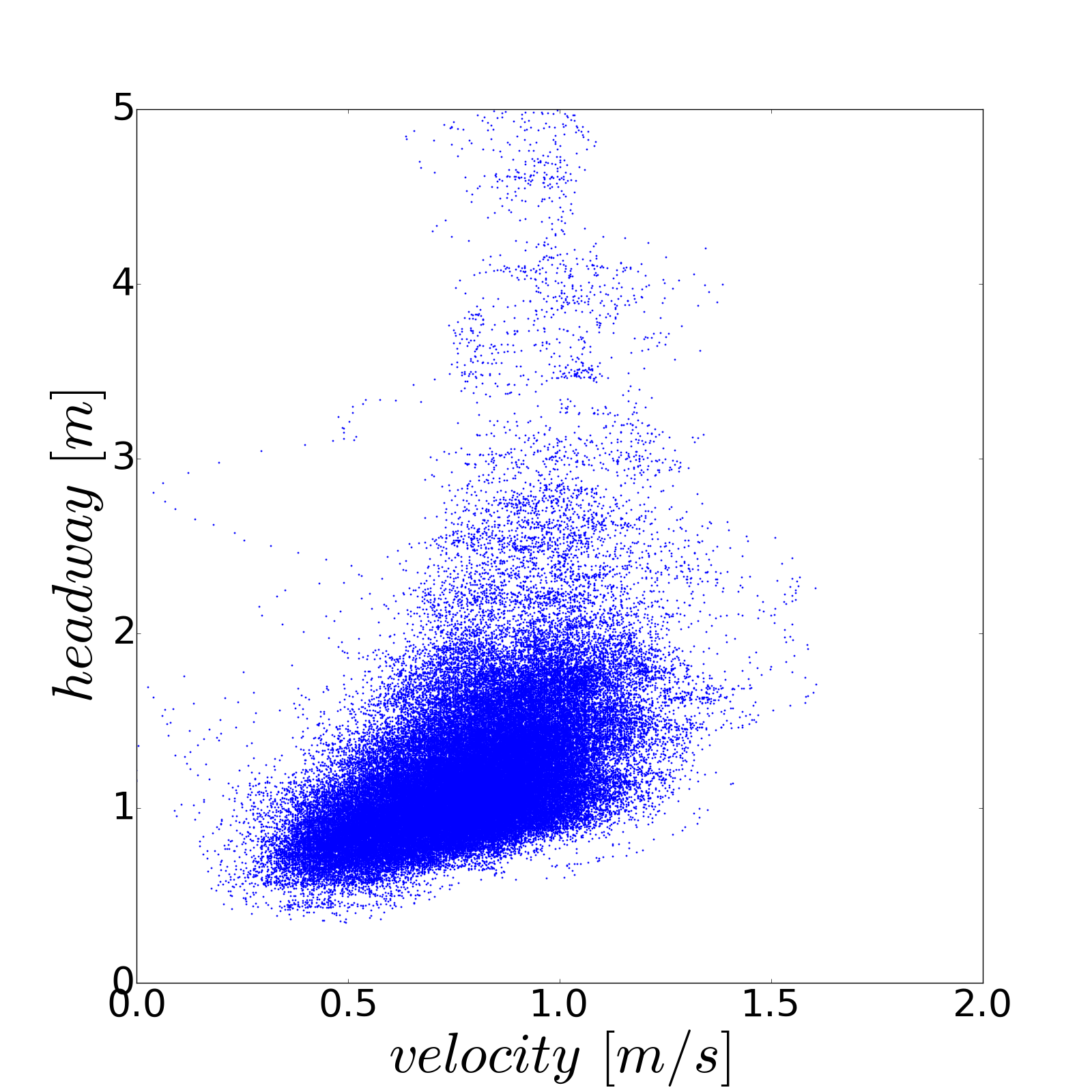}}
 \caption{\label{fig-headway-velocity-micro}The scatter diagrams of headway-velocity for different groups.}
 \end{figure*}

 From the experiment in Ref. \cite{Seyfried2005}, it is found that the individual velocity depends on spatial headway with a linear relation. Jeli\'c et al. \cite{Jelic2012a} did a similar experiment in a circle. With different experiment setup their data cover a larger range of densities especially including more data under free flow state. Surprisingly, three regimes (free regime, weakly constrained and strong constrained regimes) with different slopes between headway and velocity are observed. Fortunately, from our experimental setup we also get data with a quite large range of densities. FIG.\ref{fig-headway-velocity-micro} shows the scatter diagrams of the instantaneous individual velocity $v_i(t)$ and headway $d_{h,i}(t)$. At first sight, there are two regimes for all the three groups. One is free regime where the individual velocity is independent on the size of headway, while in the other regime it indicates a linear dependency between velocity and headway. The free velocities and the slopes of linear regime  are different for the three groups.

\begin{figure*}
 \centering\subfigure[Headway-velocity]{
 \includegraphics[width=0.4\textwidth]{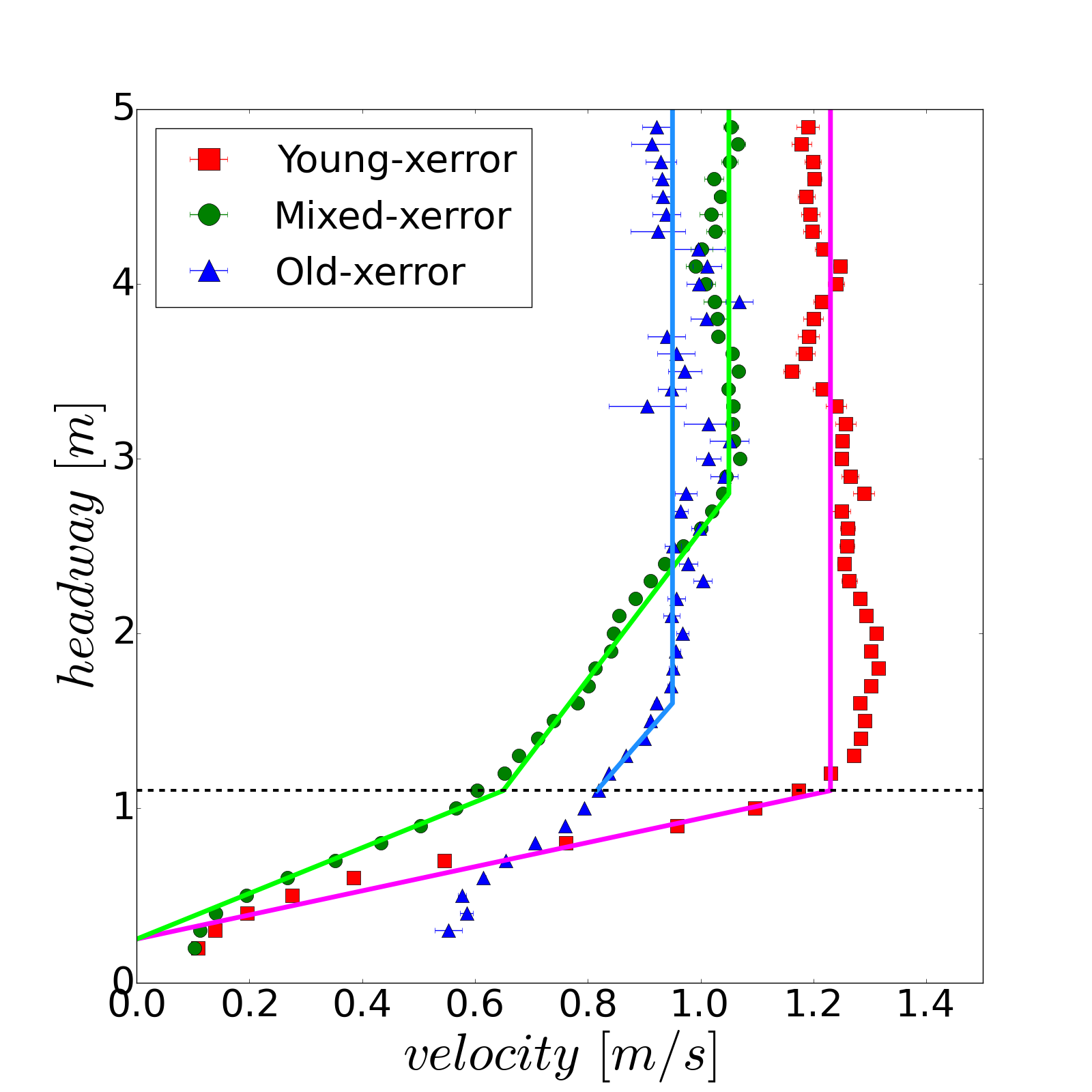}}
 \centering\subfigure[Distribution of free velocity]{
 \includegraphics[width=0.4\textwidth]{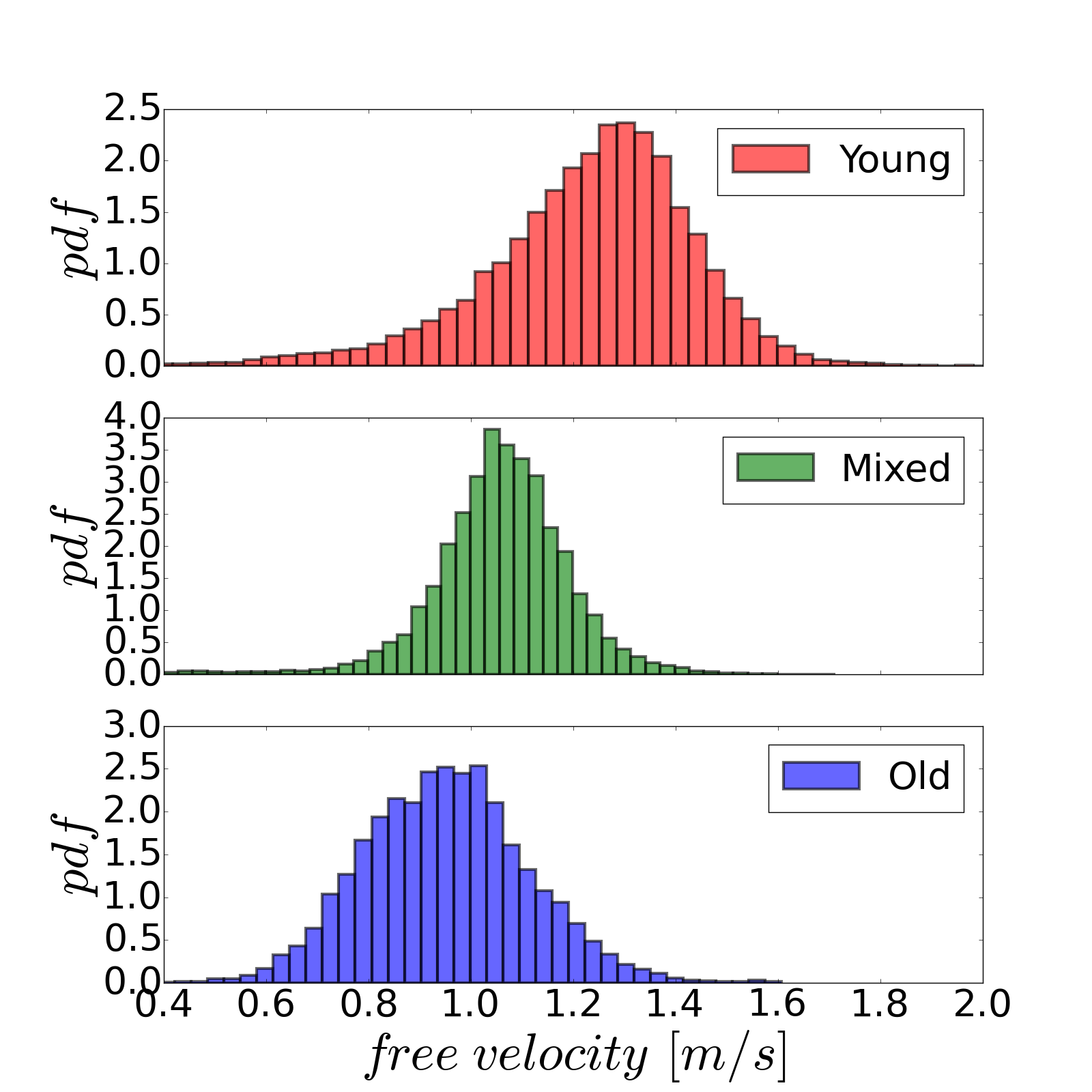}}
 \caption{\label{fig-headway-velocity-regimes}(a) Headway-velocity. The figure is obtained as the result of a horizontal binning procedure over a large number of measurements. The points are the mean values of velocity for each bin and the error bars give the standard error. The horizontal line represents headway $d_h = 1.1~ m$, and the lines with red, green and blue colors are the fitting results in different regimes for young, mixed and old group. (b) The distribution of free velocity for the three groups from the data under free regime. The mean velocities and standard deviations are $1.23 \pm 0.22~ m/s$, $1.05 \pm 0.16~ m/s$ and $0.95 \pm 0.16~m/s$ for young student, old and mixed group respectively. }
 \end{figure*}

 To gain a deeper understanding on this issue, we use the similar binning procedure as in Ref. \cite{Jelic2012a} to analyze the data and check whether the same relations between headway and velocity can be found in the young student, old and mixed group. From the individual headways and the corresponding velocities, one pair of dataset is selected per second for each pedestrian to guarantee the independency of the samples. The headway-velocity diagram is plotted using a horizontal binning procedure in FIG.\ref{fig-headway-velocity-regimes}(a). In this procedure, firstly the whole data of headway is divided into a series of bins per 0.1 $m$. Then the number of velocities falling into each bin is counted and the mean value and standard error are calculated. For lack of data in old group, here we mainly focus on the analysis of young student group and mixed group. It is found that different regimes are observed in different groups:

(a) When the headway is greater than a specific value (1.1 $m$ for young group, 1.6 $m$ for old group and 2.8 $m$ for mixed group), the velocity is independent on headway and pedestrians walk with their preferred velocities. This is the free regime as called in Ref. \cite{Jelic2012a}. The mean values of the free velocity are 1.23 $m/s$, 0.95 $m/s$ and 1.05 $m/s$ for young, old and mixed group respectively and their distributions are shown in FIG.\ref{fig-headway-velocity-regimes}(b). It is interesting that for the mixed group, only one peak is observed from FIG.\ref{fig-headway-velocity-regimes}(b), which means pedestrians with different mobility try to move in a compromised velocity and maintain the same pace with each other but not keep their original preference. In this case, an old pedestrian has to accelerate from his/her normal speed and needs more space ahead. At the same time, the young student in front tries to decelerate to avoid potential collisions during movement. Besides, the young students and old people are arranged in the corridor alternately. They belong to different age ranges and are not familiar with each other, thus prefer to keep a relatively larger distance between each other. This could be the reason why the mixed group needs more space (2.8 $m$) to reach the free velocity compared to other groups (1.1 $m$ for young group, 1.6 $m$ for old group). In addition, young students have very small age difference ($<$ 2 years) and familiar with each other very well, which leads to nearly the same mobility and homogeneous composition of crowd. At free flow state, it is easier for them to adjust their paces to reach a compromise with others in the crowd and keep certain speed.

(b) When the headway is between 1.1 $m$ and 2.8 $m$ for mixed group (between 1.1 $m$ and 1.6 $m$ for old group), the velocities of pedestrians depend weakly on the spatial headway and they do not slow down much. This regime is the weakly constrained regime. However, the weakly constrained regime is not observed in young student group.

(c) When the headway is smaller than the specific value (1.1m for both young and mixed group), it is called strongly constrained regime and the adaptation of the velocity to the spatial headway is much more pronounced. For this regime, the values of the transition headway are also consistent with the findings in Ref. \cite{Jelic2012a}. Due to the lack of data at high density case ($\rho > 0.9~m^{-1}$ corresponding to $d_h < 1.1~m$), the strongly constrained regime for the old group is not clear and needs further studies in the future.

 \begin{figure}
 \includegraphics[width=0.40\textwidth]{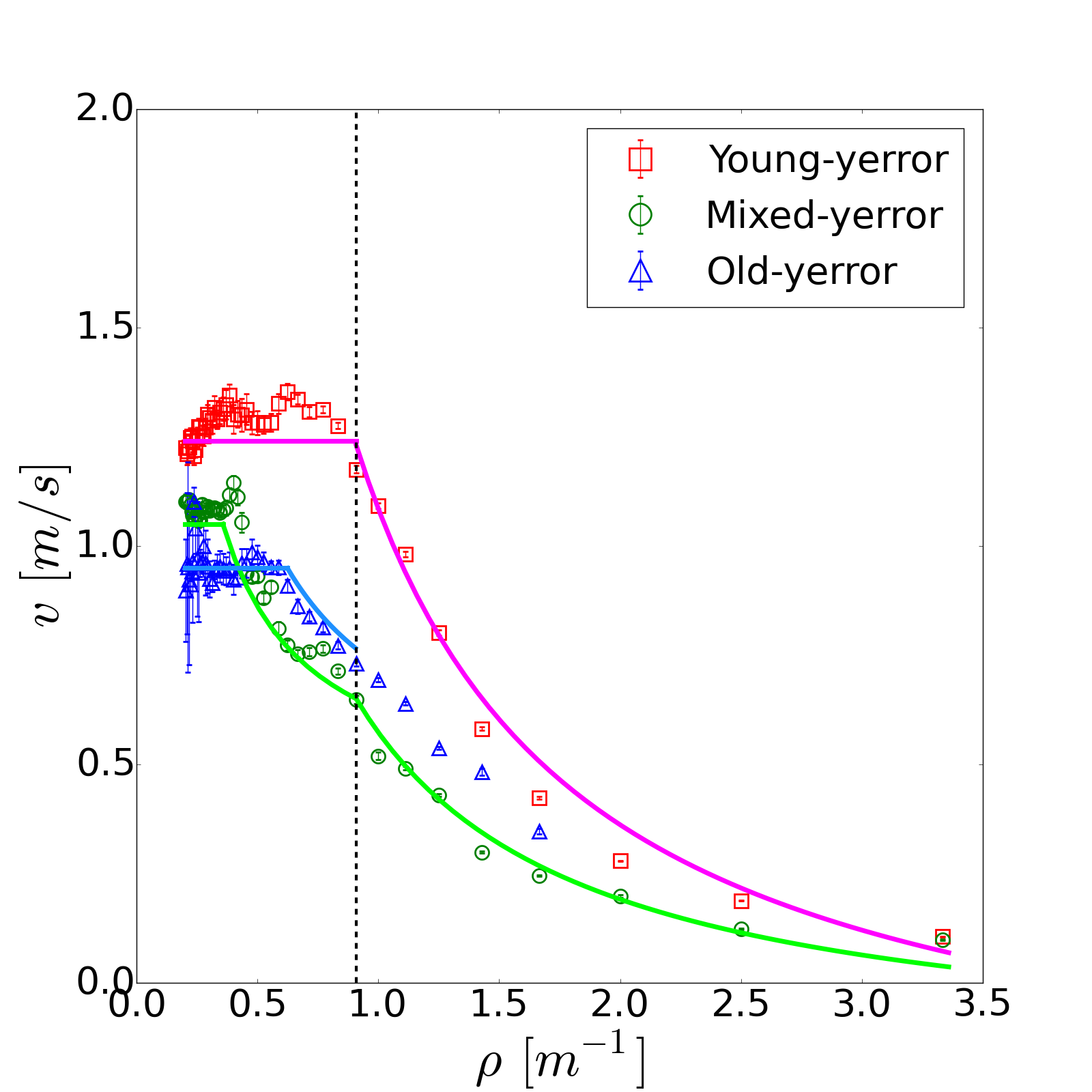}
 \caption{\label{fig-FD-binning}The density-velocity relation on macro level by using binning procedure.The bars are the binning result for data from  FIG.\ref{fig-FD-groups}. The density interval adopted in binning procedure corresponds to the headway interval used in FIG.\ref{fig-headway-velocity-regimes}(a).The curves with different colors are obtained according to the slopes and intercepts of the corresponding fitting lines in FIG.\ref{fig-headway-velocity-regimes}(a).}
 \end{figure}

 The students in the young group are more homogeneous in age composition, body dimension, educational level and empirical experience etc. However, pedestrians in the mixed group have a great difference on these aspects, which can lead to different kinds of movement behaviors and make the crowd more inhomogeneous. The different mobilities and adaptive abilities of pedestrians in the system at a microscopic level result in the discrepancies of the macroscopic movement properties of the crowd. This may be the reason why three regimes in mixed group but only two regimes in young student group are observed.

As interpreted in Ref. \cite{Seyfried2005} and \cite{Jelic2012a}, the slope in each linear regime has the dimension of time and represents the sensitivity to the spatial headway. It is called as the adaptation time that available to react before being at the minimal distance from the current predecessor \cite{Jelic2012a}. In TABLE \ref{Table_adaption_time} we compare the adaption time obtained from our experiment with that from France in weakly and strongly constrained regimes. For young student group the weakly constrained regime is not observed in our experiment, whereas the adaption time in strongly constrained regime for old group is not obtained for the lack of experimental data. In weakly constrained regime, both mixed group and old group in our experiment have longer adaption time than pedestrians in France. The old group is more sensitive to the spatial headway than the young student group. However, different results are observed in strongly constrained regime. The mixed group has the longest adaptation time 1.31 $s$, whereas the young student group has the shortest time (0.69 $s$) which is close to the time (0.74 $s$) in the France experiment. It seems that the system with mixed group becomes more complicated and pedestrians need more time to adapt their movement.

%For $\rho < 0.9~m^{-1}$, the flow increases with a unique slope for young student group but different slopes for other two groups.
%These differences can be also found from the fundamental diagram in FIG.\ref{fig-FD-groups}.
Macroscopically, the slopes of these fitting lines in FIG.\ref{fig-headway-velocity-regimes}(a) are related to the decay rates of the fundamental diagram in FIG.\ref{fig-FD-groups}. In FIG.\ref{fig-FD-binning}, the binning procedure is applied to the fundamental diagram. By comparing the binning result with the curves obtained according to the slopes and intercepts of the fitting lines in  FIG.\ref{fig-headway-velocity-regimes}(a), different regimes obtained on micro level (FIG.\ref{fig-headway-velocity-regimes}(a)) are also found on macro level (FIG.\ref{fig-FD-binning}). Therefore, the existence of different regimes is verified on both micro and macro levels.

\begin{table}%[H] add [H] placement to break table across pages
 \caption{\label{Table_adaption_time}Comparison of the adaptation time between our experiment and France experiment. The value for the free regime is not meaningful \cite{Jelic2012a}, thus we do not display and compare the values in this regime.}
 \begin{ruledtabular}
 \begin{tabular}{cccccc}
 &\multicolumn{3}{c}{China}&\\[-2mm]
Regime	& \multicolumn{3}{c}{\rule{3.5cm}{0.013cm}}&France\\
&Young &Old &Mixed&\\
\hline
Free	&-&-&-&13.7\\
Weakly constrained&-&3.79&4.25&5.32\\
Strongly constrained&0.69&-&1.31&0.74\\
 \end{tabular}
 \end{ruledtabular}
 \end{table}

\begin{table}%[H] add [H] placement to break table across pages
 \caption{\label{Table_slope}Comparison of the intercept and slope in our experiment and others in different countries \cite{Jelic2012a}.}
 \begin{ruledtabular}
 \begin{tabular}{cccccccc}
 &\multicolumn{3}{c}{China}&\\[-2mm]
Regime	& \multicolumn{3}{c}{\rule{3cm}{0.013cm}}&France&Germany&India\\
&Young &Old &Mixed&\\
\hline
Intercept ($m$)&0.25&-&0.25&0.45&0.37(0.36)&(0.22)\\
Slope ($s$)	&0.69	&-	&1.31	&0.75	&1.01(1.04)&(0.89)\\

 \end{tabular}
 \end{ruledtabular}
 \end{table}

\begin{figure}
 \includegraphics[width=0.40\textwidth]{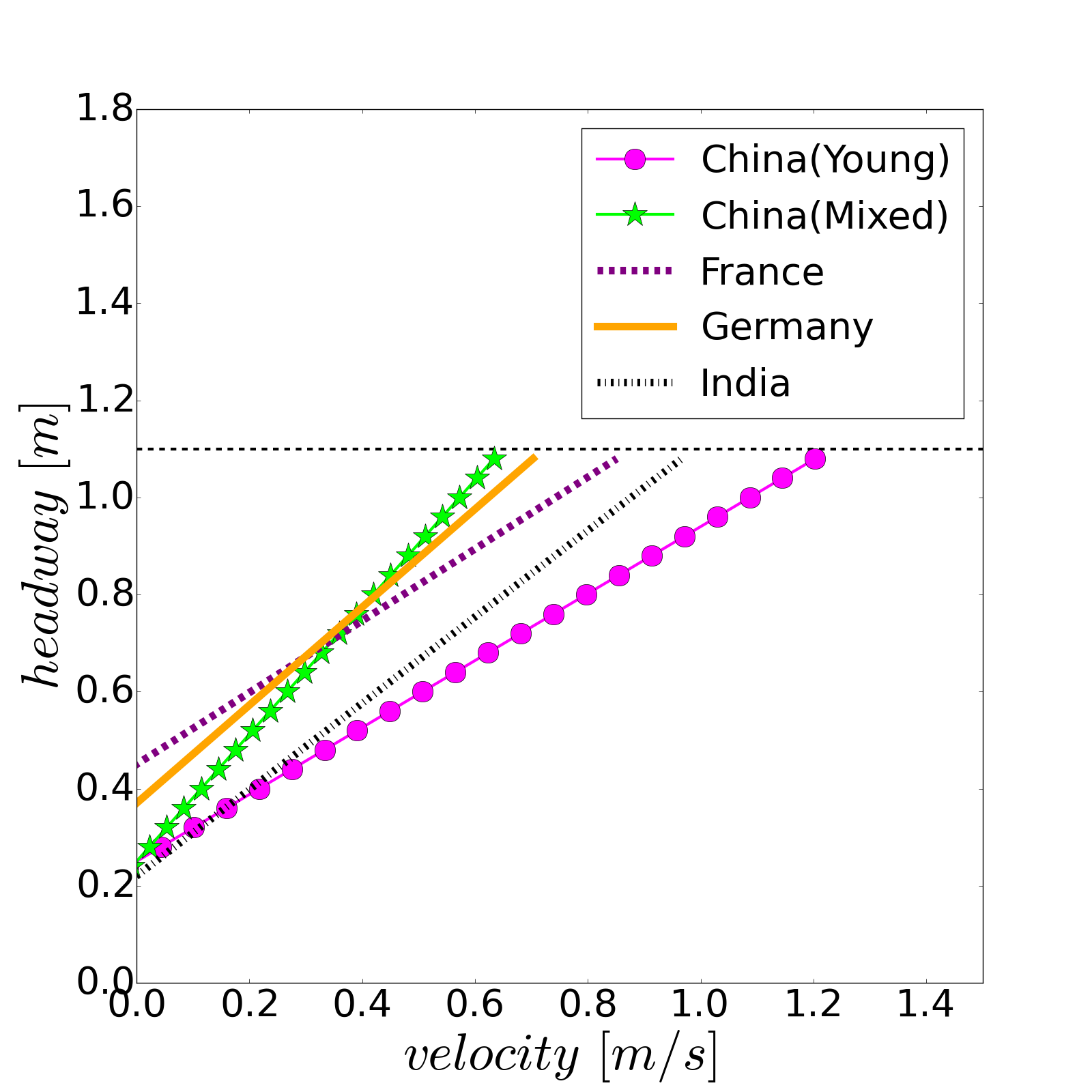}
 \caption{\label{fig-slope}The linear fit results of headway-velocity in the strongly constrained regime for experiments from different countries.  Our data are shown in line with symbols, while others are illustrated only with lines. The horizontal line represents $d_h = 1.1~m$.}
 \end{figure}

Besides, we compare the slope and intercept with the zero velocity axis of the linear fit to the headway-velocity relationship in the strongly constrained regime for the experiment from different countries.  As shown in TABLE \ref{Table_slope} and FIG.\ref{fig-slope}, the intercept (0.25 $m$) from our data is close to that from India but smaller than that from Germany and France experiments. The intercept can be seen as a minimal personal space. On one hand, the average body size of Asian is smaller compared to European. On the other side, it seems that Chinese and Indians are rather to accept small distance and contact with neighbors at high density situations, whereas French pedestrians prefer to break hard and German pedestrians try to walk at lower velocity. However, the slopes are different for all the groups from different countries.

 Although there seems to be a kind of universality in pedestrian's behavior at high density situation, it is obvious that differences exist among different experimental results. Lots of potential factors may lead to such discrepancies. For example, the experimental conditions in these experiments are different. The geometry in our experiment  includes two straight corridor and two semicircles, while it is an exact circle in Ref. \cite{Jelic2012a}. The circumferences of the geometries are also different. However, the composition of the test pedestrians may be the main reason. Pedestrians in our study are young students (mean age of 17) and old people (mean age of 52), while pedestrians in other experiments are university students. Different educational and cultural backgrounds may also do influence their behaviors.

\section{Conclusion}

The effect of different compositions of crowd (young student group, old people group and mixed group) on pedestrian dynamics is investigated by a series of single-file experiments. All the analyses in the paper are based on the pedestrian trajectories extracted by the software $PeTrack$. Different free velocities (1.23 $m/s$, 0.95 $m/s$ and 1.05 $m/s$) are observed for young, old and mixed group at low density respectively. From the time-space diagram, stop-and-go waves are observed and they propagate in the opposite direction to the movement of pedestrians with velocities roughly 0.4 $m/s$ for $\rho_g \approx 1.75 ~m^{-1}$ and 0.3 $m/s$ for $\rho_g > 2.3~ m^{-1}$. It occurs more easily and frequently in mixed group, which indicates density is not the only reason for the emergence of stop-and-go waves.

The fundamental diagrams for the three groups are totally different. The flow of young group is the largest at the same density among three groups. However, the maximal flows are observed at nearly the same density $\rho_g = 0.9~ m^{-1}$. Throughout nondimensionalization by considering the free velocity and body size, the difference among the three fundamental diagrams seems smaller and they can agree well at the beginning of free flow state and congested state with stop-and-go wave. But at medium density situation, obvious discrepancies can be observed. In the mixed group, young students and old people influence each other, which causes the movement properties of the mixed groups are quite complex and are not able to be obtained from other groups by simply scaling from free velocity and body dimension.

From the analysis of velocity-headway relation, three linear regimes (free, weakly constrained and strongly constrained regime) are observed in the mixed group, which is consistent with the findings in Ref. \cite{Jelic2012a}. However, only two linear regimes are observed in the young student group. The slopes and intercepts with the zero velocity axis are also different for the data from different countries. Facing congestion, pedestrians from different groups will take different strategies. The influence of the educational and cultural background as well as the composition of the crowd on the properties of pedestrian movement should not be ignored.

However, as for the validation of the non-dimensional method and headway-velocity relation, more experimental data are still needed especially in higher density range for old people.

\begin{acknowledgments}
The foundation supports from the State Key Laboratory of Fire Science in University of Science and Technology of China (HZ2015-KF11) and the China Scholarship Council (CSC).
\end{acknowledgments}

% Create the reference section using BibTeX:
\bibliography{basename of .bib file}

\end{document}